\documentclass{aa}  

\usepackage{txfonts}
\def\ebv {$E(B-V)$}
\def \GALEX {GALEX}
\def \Herschel {\textit{Herschel}}
\def \Planck {\textit{Planck}}

\begin{document}

   \title{The \GALEX{} Ultraviolet Virgo Cluster Survey (GUViCS). V.}

   \subtitle{Ultraviolet diffuse emission and cirrus properties in the Virgo cluster direction}

   \author{S. Boissier\inst{1}, A. Boselli\inst{1}, E. Voyer\inst{1}
          \and
S. Bianchi\inst{2} \and C. Pappalardo\inst{3,4} \and P. Guhathakurta\inst{5}
\and
S. Heinis\inst{6}, L. Cortese\inst{7}
\and
P.-A. Duc\inst{8}, J.-C. Cuillandre\inst{8}
\and
J. I. Davies\inst{9}, M. W. L. Smith\inst{9} 
}
\institute{Aix Marseille Universit\'e, CNRS, LAM (Laboratoire d'Astrophysique de Marseille) UMR 7326, 13388, Marseille, France\\
             \email{firstname.name@lam.fr}
\and
INAF-Osservatorio Astrofisico di Arcetri, Largo Enrico Fermi 5, I-50125, Firenze, italy  
\and
Centro de Astronomia e Astrof\'{\i}sica da Universidade de Lisboa,
Observat\'{o}rio Astron\'{o}mico de Lisboa, Tapada da Ajuda, 1349-018 Lisboa, Portugal
\and 
Instituto de Astrof\'{\i}sica e Ciencias do Espa\c{c}o, Universidade de Lisboa, OAL, Tapada da Ajuda, PT1349-018 Lisboa, Portugal 
\and
University of California Observatories/Lick Observatory, University of California Santa Cruz, Department of Astronomy \& Astrophysics, 1156 High Street, Santa Cruz, California 95064, USA
\and
Department of Astronomy, University of Maryland, Stadium Drive, College Park, MD 20742-2421, USA
\and
Centre for Astrophysics \& Supercomputing, Swinburne University of Technology, Mail H29 -- PO Box 218, Hawthorn   VIC   3122, Australia 
\and
Laboratoire AIM, Service d'astrophysique, CEA-Saclay, Orme des merisiers, b\^atiment 709, 91191 Gif sur Yvette cedex, France
         \and
School of Physics and Astronomy, Cardiff University, The Parade, Cardiff, CF24 3AA, UK  
   }
   \date{accepted in A\&A, april 2015}
  \abstract
   {The Virgo direction has been observed at many wavelengths in the recent years, in particular in the ultraviolet with \GALEX{}. The far ultraviolet (FUV) diffuse light detected by \GALEX{} bears interesting information on the large scale distribution of Galactic dust, 
owing to the \GALEX{} FUV band sensitivity and resolution.}
   {We aim to characterise the ultraviolet large scale distribution of diffuse emission in the Virgo direction. A map of this emission may become useful for various studies by identifying regions where dust affects observations by either scattering light or absorbing radiation.}
   {We construct mosaics of the FUV and near ultraviolet diffuse emission over a large sky region (RA 12 to 13 hours, DEC 0 to 20 degrees) surrounding the Virgo cluster, using all the \GALEX{} available data in the area. We test for the first time  the utilisation of the FUV diffuse light as a Galactic extinction \ebv{} tracer.}
   {The FUV diffuse light scattered on cirrus reveals details in their geometry. Despite a large dispersion, the FUV diffuse light correlates roughly with other Galactic dust tracers (coming from IRAS, \Herschel, \Planck), offering an opportunity to use the FUV emission to locate them in future studies with a better resolution (about 5 arcsec native resolution, 20 arcsec pixels maps presented in this paper) than several usual  tracers. Estimating the Galactic dust extinction on the basis of this emission allows us to find a smaller dispersion in the $NUV-i$ colour of background galaxies at a given \ebv{} than with other tracers. The diffuse light mosaics obtained in this work are made publicly available.\thanks{The FUV and NUV diffuse light mosaics will be available on the 
GUViCS web site http://galex.lam.fr/guvics/mosaics.html and at the CDS via anonymous ftp to cdsarc.u-strasbg.fr (130.79.128.5)
or via http://cdsweb.u-strasbg.fr/cgi-bin/qcat?J/A+A/ }
}
   {}
   \keywords{dust, extinction, Ultraviolet: ISM,Ultraviolet: galaxies}
 \titlerunning{The GUViCS FUV diffuse emission}
\authorrunning{Boissier et al.}
   \maketitle
%

\section{Introduction}

Large scale diffuse structures have been found early-on in the Galaxy \citep{devauc55}, 
\citet{sandage76} publishing for the first time detailed optical images of high latitude clouds found in a Galactic polar caps survey.
However, It is only with the IRAS satellite that we were able to discover a rich geometry of far infrared emitting cirrus clouds  over the full sky \citep{low84}. Deep optical imaging then revealed their emission due to dust-scattering, showing a good correspondence with the far infrared \citep[e.g.][]{devries85,guha94,witt2008}. 
The cirrus are also seen in the far ultraviolet (FUV) domain, in which a major component of 
the light results from dust scattering \citep{witt97}.

Early studies of the FUV diffuse light can be found in
\citet{lillie76,paresce1980,jakobsen1984,murthy89,fix89,hurwitz91,perault91,murthy99}.
\citet{haikala95} discovered with the FAUST telescope Galactic cirrus directly from their FUV emission.
With the \GALEX{} telescope, and its All sky Imaging Survey (AIS), extensive studies  of the diffuse FUV emission became possible \citep[e.g.][]{sujatha10,hamden2013,murthy14}. 

Studying the cirrus in FUV is useful for many reasons:
\begin{itemize}
\item UV data allows
the determination of the scattering dust properties in this wavelength domain, constraining dust grain properties 
\citep[e.g.][]{witt97,sujatha05}.

\item Recently,  \citet{seon14} and \citet{hodges2014} have found ultraviolet haloes in edge-on galaxies that they interpret as reflection nebula from dust outside galaxies. While it is a different component than the cirrus themselves, characterising the dust properties in our backyard is also useful for such extra-galactic studies, provided the dust is of similar nature.

\item Several works found an excess red emission (between 6000 and 8000 $\AA$) in reflection nebulae, resulting from a luminescence excited by the FUV radiation of illuminating stars \citep{witt85}. \citet{guha89,guha94,gordon98,szomoru98,witt2008} have shown that such Extended Red Emission (ERE) is also found in the diffuse interstellar medium and cirrus clouds.

\item The presence of cirrus in our Galaxy directly affects the colour or detection of objects in deep observations of extra-galactic sources, since cirrus modify the background to which the objects are contrasted. It is especially true for low surface brightness regions in nearby galaxies such as tidal features that may be hard to distinguish from Galactic dust emission, or for lensing studies. This point is well illustrated by the Fig. 10 of \citet{duc15}.

\item The presence of Galactic cirrus also affects UV/optical studies of extra-galactic sources via dust extinction \citep[e.g.][]{burstein82}. Usually, data are simply corrected using \citet{schlegel98} maps based on the 100 microns dust emission.
\end{itemize}

Recently, a large area around the Virgo cluster has been surveyed at many wavelengths: especially by \GALEX{} in the FUV and near ultraviolet (NUV) bands forming the GUViCS survey \citep{boselli2011}. 
These data have been used already for several works studying 
the effect of the cluster environment on the evolution of galaxies \citep{cortese11,boselli14}, 
the effect of ram-pressure on star formation \citep{boissier12}, 
or stripped systems \citep{arrigoni12}. The source catalogs have been publised in \citet{voyer2014}.

The area has also been observed in the frameworks of the Next Generation Virgo Cluster Survey (NGVS) in the optical \citep{ferrarese2012}, of the \Herschel{} Virgo Cluster Survey
(HeViCS) in the far infrared \citep{davies2010}, and of the Arecibo Legacy Fast ALFA survey (ALFALFA) in HI
\citep{giovanelli2005}. Deep B and V images (reaching 28.5 mag arcsec$^{-2}$ in the B band) 
were obtained by \citet{mihosinprep} in about 15 deg$^2$ around M87 and M89, using the Schmidt
telescope at Kitt Peak.
While several of these surveys are mostly dedicated to extra-galactic science, they can also be used to study the cirrus component in our Galaxy, or may be affected by them.

In this paper, we concentrate on the UV data from the GUViCS survey. We characterise the cirrus in the UV and compare our findings to other wavelengths. 
Recently, \citet{hamden2013} performed a full sky study of the ultraviolet diffuse light in the \GALEX{} All sky Imaging Survey.
We focus instead on the Virgo cluster area, taking advantage of the large number of observations performed allowing us to use deeper data, and
to visually inspect FUV detected cirrus and compare them to ancillary data in a small region of the sky, while their work is statistical in nature.
This will be useful to any analysis based on deep data (from the many recent surveys of Virgo) that may be subject to pollution by diffuse low surface brightness emission. For instance, our FUV maps were already used by \citet{durrell2014} to make sure their results concerning the 
spatial distribution of globular clusters in Virgo were not affected by cirrus.
The presence of cirrus may be crucial in the study of low surface brightness features that may be associated  to tidal interactions, expected to be frequent in the cluster environment of Virgo. A school case is the interacting pair  NGC 4435/4438.  
\citet{cortese2010} studied this system with a combination of multi-wavelength data including far infrared and far ultraviolet observations.
These data helped them to elucidate the nature of  a  ``plume'' detected in deep optical data.
The cirrus are also easily recognized in a simple visual examination of the deep optical imaging of \citet{mihosinprep}. While they seem to correspond well to the location where we detect them in FUV or infrared, a detailed analysis (distinguishing the cirrus vs faint emission linked to the galaxies, and comparing to other wavelengths) will be performed in the future.

In Sect. \ref{secMos}, we present our data and how we constructed the mosaics (made publicly available with this paper) 
over a large sky area (RA 12 to 13 hours, DEC 0 to 20 degrees) surrounding the Virgo cluster. An analysis (UV properties of cirrus, comparison with other wavelengths) is performed in Sect. \ref{SecAna}, and our conclusions are summarised in Sect. \ref{SecConclu}

   \begin{figure}
   \centering  
   \includegraphics[width=10.5cm]{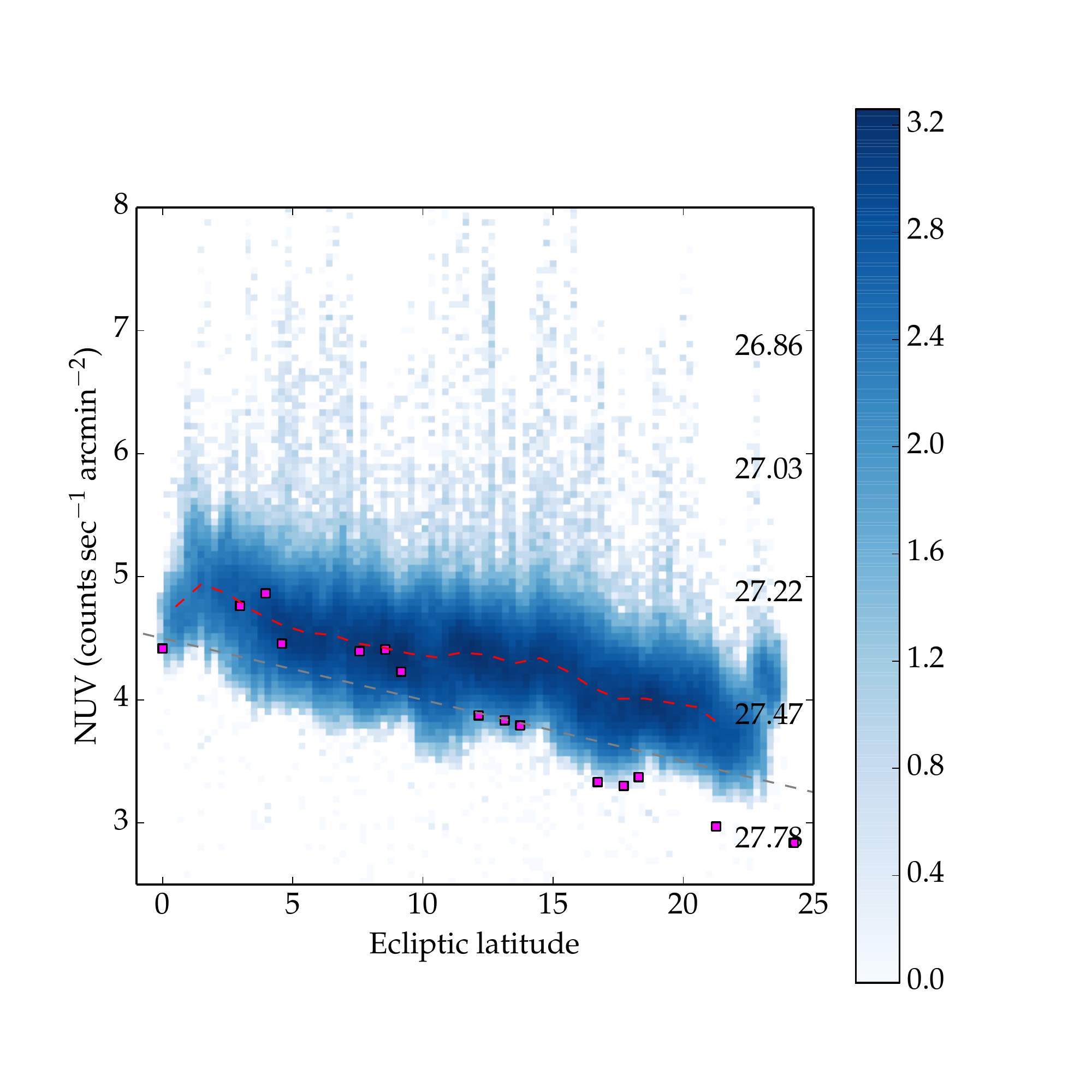}
   \caption{Objects-subtracted NUV counts (based on 1 arcmin pixels, median-filtered in 3x3 pixels boxes) as a function of the ecliptic latitude. The numbers on the right axis indicate the surface brightness in units of AB magnitude per square arcsec. The colorbar indicates the decimal logarithm of the number of pixels per 2D bin. The red line shows the average. The grey line is the adopted background (Sect. \ref{secdefinenuvbg}). The magenta squares report predictions of the \GALEX{} zodiacal light tool.}
     \label{FigBackvsLatNUV}%
    \end{figure}

   \begin{figure}
   \centering  

    \includegraphics[width=10.5cm]{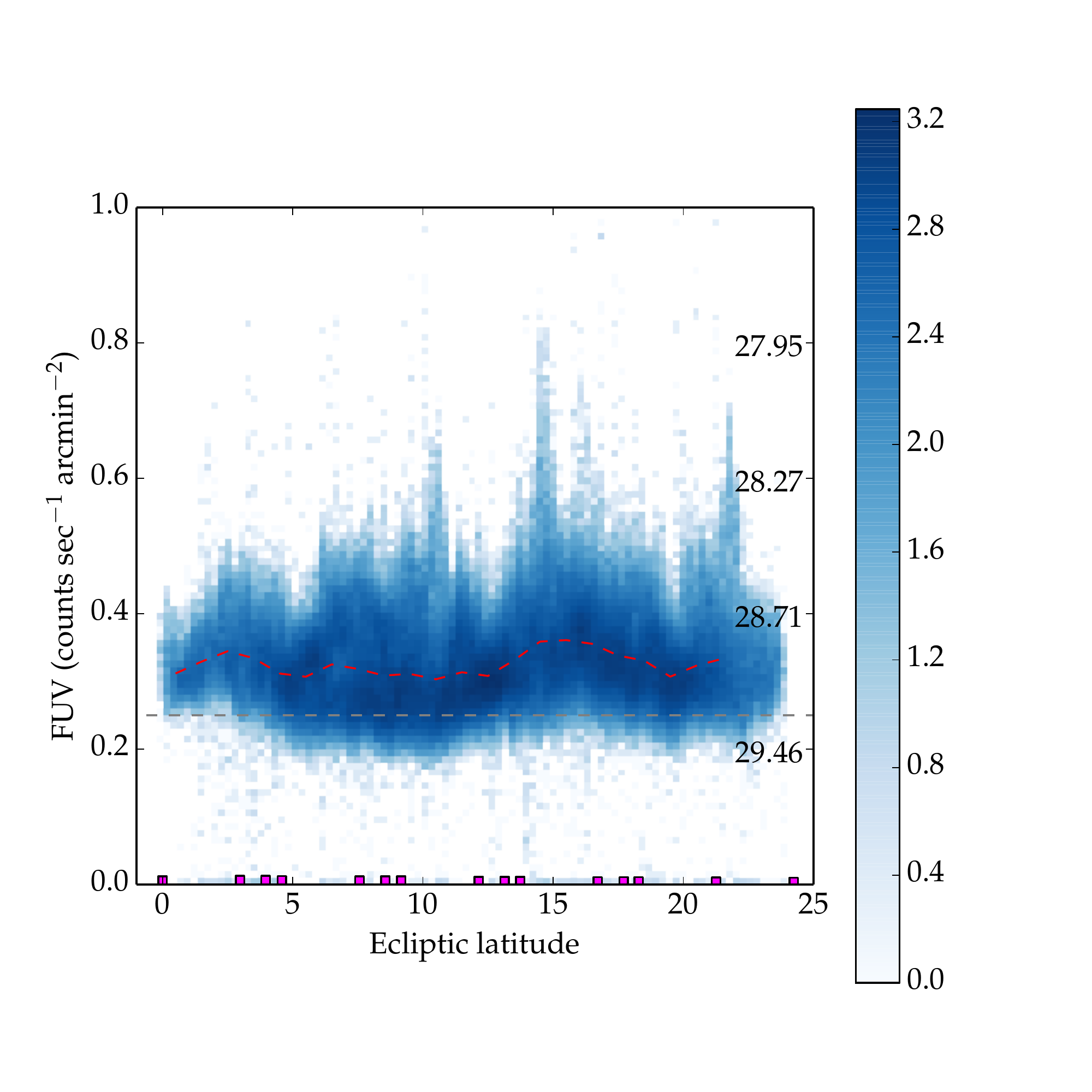} 
   \caption{Objects-subtracted FUV counts (based on 1 arcmin pixels, median filtered in 3x3 pixels boxes) as a function of the ecliptic latitude. The numbers on the right axis indicate the surface brightness in units of AB magnitude per square arcsec.The colorbar indicates the decimal logarithm of the number of pixels per 2D bin. The red line shows the average. The grey line is the adopted background (Sect. \ref{secfuvbg}). The magenta squares report predictions of the \GALEX{} zodiacal light tool.}
     \label{FigBackvsLatFUV}%
    \end{figure}

\section{The GUViCS mosaics}
\label{secMos}

\subsection{FUV and NUV \GALEX{} mosaics}

The GUViCS survey was presented in \citet{boselli2011}. It combines a large numbers of tiles observed with \GALEX{} in the
FUV and NUV bands (with effective wavelengths respectively 1528 and 2271 \AA, and a spatial resolution of about 5 arcsecs). The minimal exposure time observed corresponds to the AIS survey (exposure times of typically 100 seconds, reaching a surface brightness of about 26 mag arcsec$^{-2}$) 
but are often typical of the Medium Imaging Survey or Nearby Galaxy Survey (1500 seconds exposure times, 28 mag arcsec$^{-2}$). A few tiles have even much deeper data because they were observed for other purposes and are included in GUViCS.
An update with all the tiles we include in our work  and several source catalogues can be found in \citet{voyer2014}.
The circular $\sim$ 1.2 degree diameter field of \GALEX{} offers a good basis to map large-scale structures. However, Galactic cirrus extend beyond this size. To circumvent this difficulty, we took advantage of having access to all the files needed to generate mosaics of the full area of the GUViCS survey in both the FUV and NUV band of the \GALEX{} telescope.

We started from the following \GALEX{} pipeline products: intensity maps, exposure maps, object detection masks.
After some tests, we decided to truncate them at a radius of 0.58 degrees 
to avoid problems often observed on the edge of \GALEX{} fields (i.e. increased noise, elongated point-spread function). 
We also removed a few tiles originally in the survey but with obvious positioning issues \citep[see][for examples]{voyer2014}. 

As a first step, in all the tiles, we replaced pixels in the object detection maps by surrounding values. 
Then, we used the Montage software\footnote{\url{http://montage.ipac.caltech.edu/}}  to co-add the objects subtracted intensity maps by weighting each image by its exposure time.
Montage allows to project the final image on an arbitrary pixel grid. We tested several resolutions looking for a balance between resolution and signal to noise. 
We present in this paper maps a mosaic with 1 arcmin size pixels over the full Virgo area for tiles from the AIS (having the advantage of being homogeneous and covering the largest possible area). We also produce an image with pixels of 20 arcsecs size showing finer details in the background structure owing to deeper exposures (i.e. exposure times of at least the 1500 seconds typical of the \GALEX{}  Medium Imaging Survey).

This method is similar to the one used in \citet{hamden2013}. The main differences are : they rely 
on sky background  images from the \GALEX{} pipeline while we replace the detected objects within the masks by 
neighbouring pixel values by ourselves (the pipeline background is smoothed and may not reproduce small-scale variations in the background). Second: they grow the masks of detected objects to limit the pollution of data by actual objects, while we keep the masks 
at their size to increase our chance to see real structures (even if the absolute flux level may be in some case polluted). 
Finally, their final map has pixels covering 11.79 arcmin$^2$, while we choose 1 arcmin$^2$ and 20$\times$20 arcsec$^2$ 
pixels in our study of a more limited area. The higher resolution allows us to see details of the structures 
revealed in the FUV with deeper data than the All Sky Imaging Survey that they use in their full sky statistical study.

The FUV mosaics obtained in this way presented a few artefacts due to the absence of data in a small area on the edge of the FUV detector. We edited manually the FUV mosaics to interpolate over such regions. During this visual inspection, we also manually masked a few sources that had been badly subtracted (usually very bright stars or extended galaxies). These corrections concern a small number of pixels. We checked that they do not affect the conclusions presented here (we did perform our analysis with both the raw mosaic and the edited image). The only visible difference is a tail of FUV bright pixels corresponding to badly subtracted objects in the uncorrected map.

As a final step, to obtain smoother maps, we found useful to median-filter our mosaics in 3 $\times$ 3 pixel boxes. This alleviates potential difficulties linked to the choice of the pixel grid origin and it reduces the noise at the very low surface brightness of the cirrus, at the price of degrading further the resolution of our mosaic to about 3 arcmin (for the lower resolution mosaic). This is the same order of magnitude as the resolution of other dust tracers to which we will compare our map.

   \begin{figure*}
   \centering  
\includegraphics[height=10.cm]{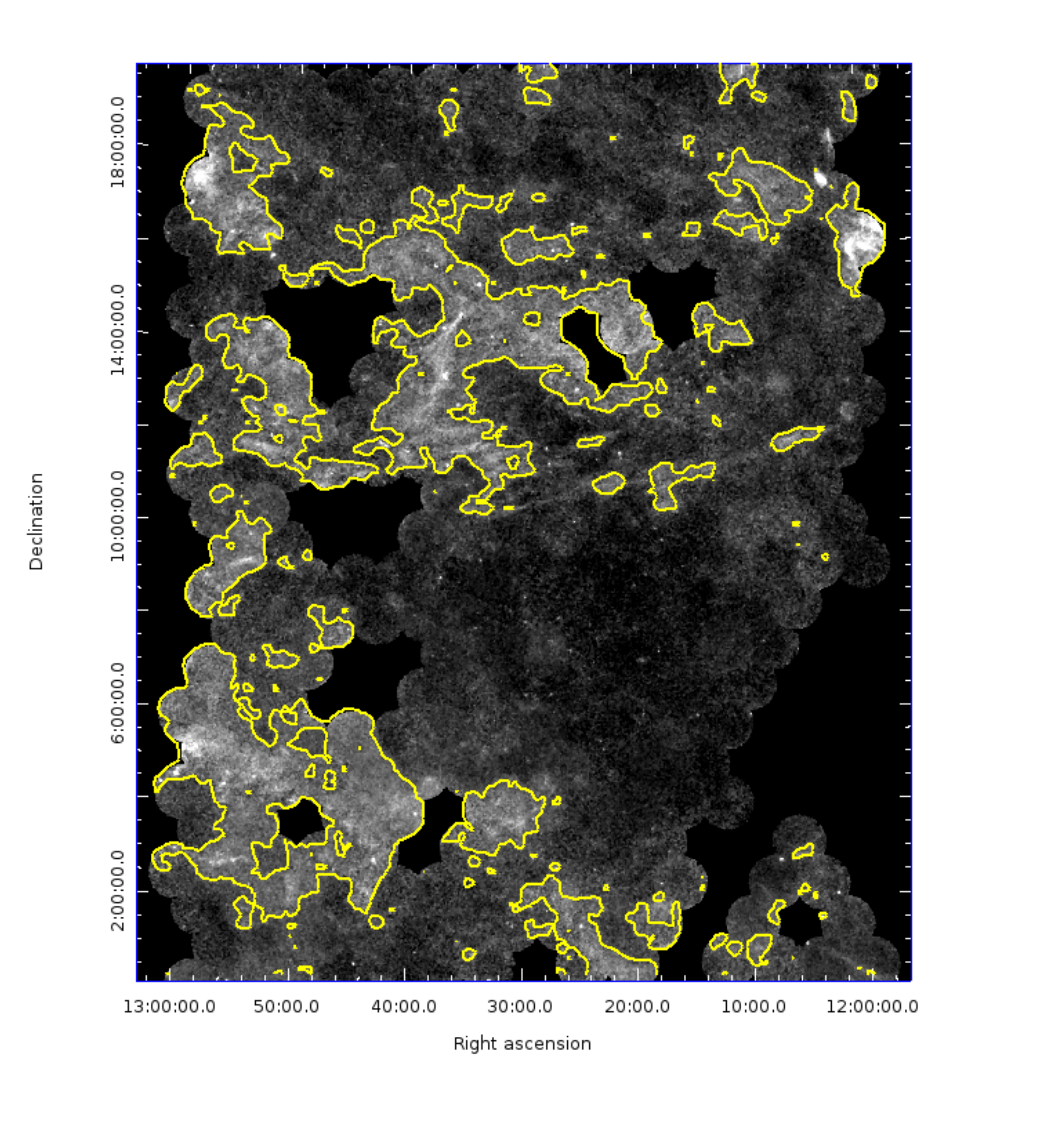}
\includegraphics[height=10.cm]{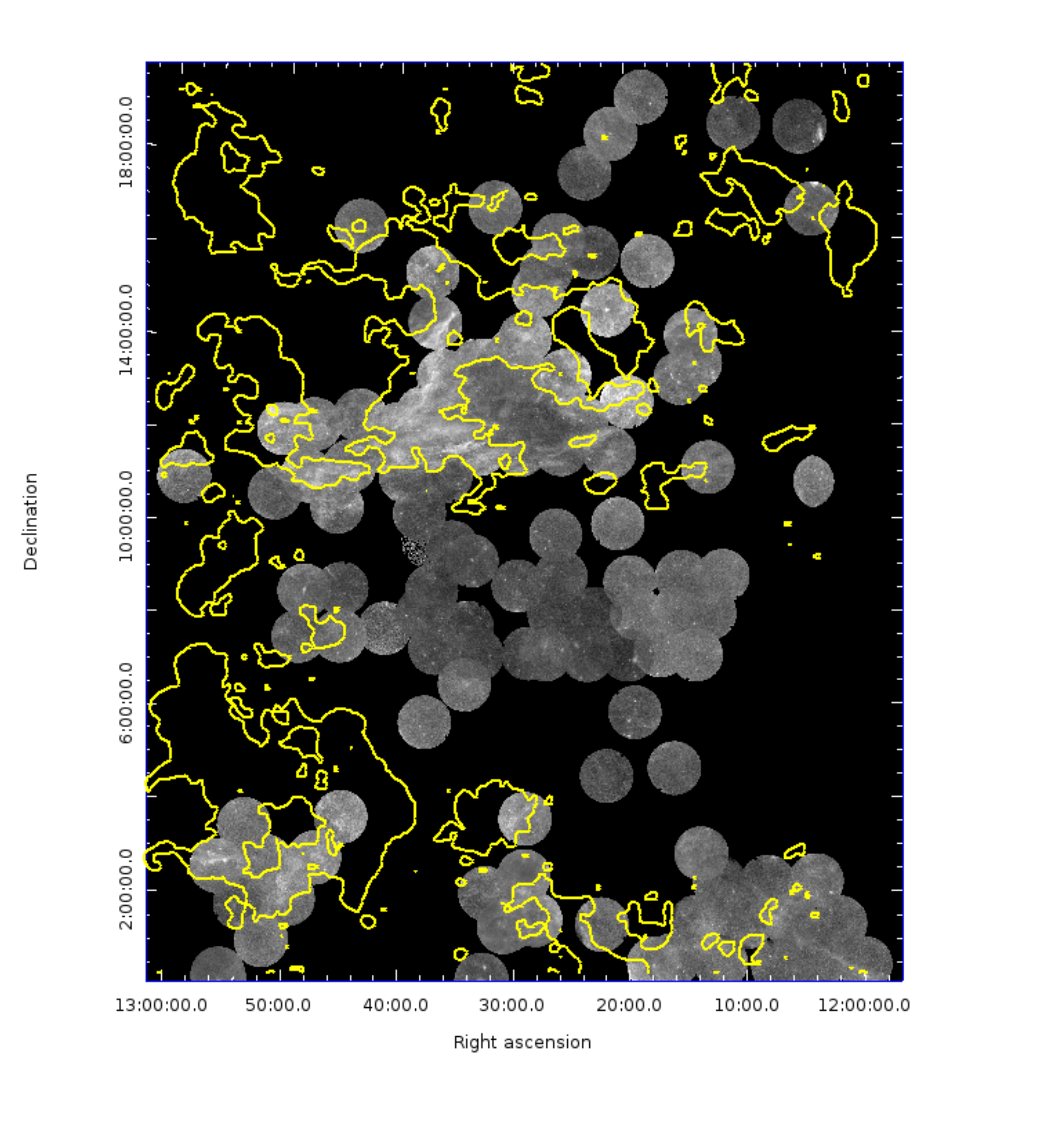}
  \caption{Left: mosaic showing the diffuse FUV emission obtained from AIS data (1 arcmin pixels, median smoothed on 3x3 pixels). Right: diffuse FUV emission obtained on the basis of deeper data (on 20 arcsec pixels, median smoothed on 3x3 pixels). The contours indicate roughly the regions in which the cirrus emission is found (Sect. \ref{sectioncontours})} 
     \label{FigFuvMaps}%
    \end{figure*}

\subsection{Units}

Different communities express the UV surface brightness in different units \citep[see][ for an exhaustive list of surface brightness units and their conversions]{leinert98}. To make it easy to convert between various units used in this paper, we express most of our maps in \GALEX{} counts (per sec). To have graspable numbers, we used counts per arcmin$^2$ in most cases. When FUV/NUV comparison was involved, we switched to AB magnitudes and surface brightnesses which are relatively standard. Conversion between counts and surface brightness is straightforward\footnote{\url{http://galexgi.gsfc.nasa.gov/docs/galex/FAQ/counts_background.html}}:
\begin{equation}
FUV {\rm \, AB \, magnitude} = -2.5 log10 ({\rm counts \, sec^{-1}}) + 18.82
\end{equation}
\begin{equation}
NUV {\rm \, AB \, magnitude} = -2.5 log10 ({\rm counts \, sec^{-1}}) + 20.08
\end{equation}

"Continuum Units" (CUs, photons cm$^{-2}$ s$^{-1}$ sr$^{-1}$ \AA$^{-1}$) are often used in studies of the FUV diffuse light. It is straightforward to convert from counts to CUs for the FUV \GALEX{} band:
\begin{equation}
{\rm counts \, sec^{-1} arcmin^{-1}} = {\rm CUs} / 1260 
\end{equation}

   \begin{figure*}
   \centering  
\includegraphics[height=9.cm]{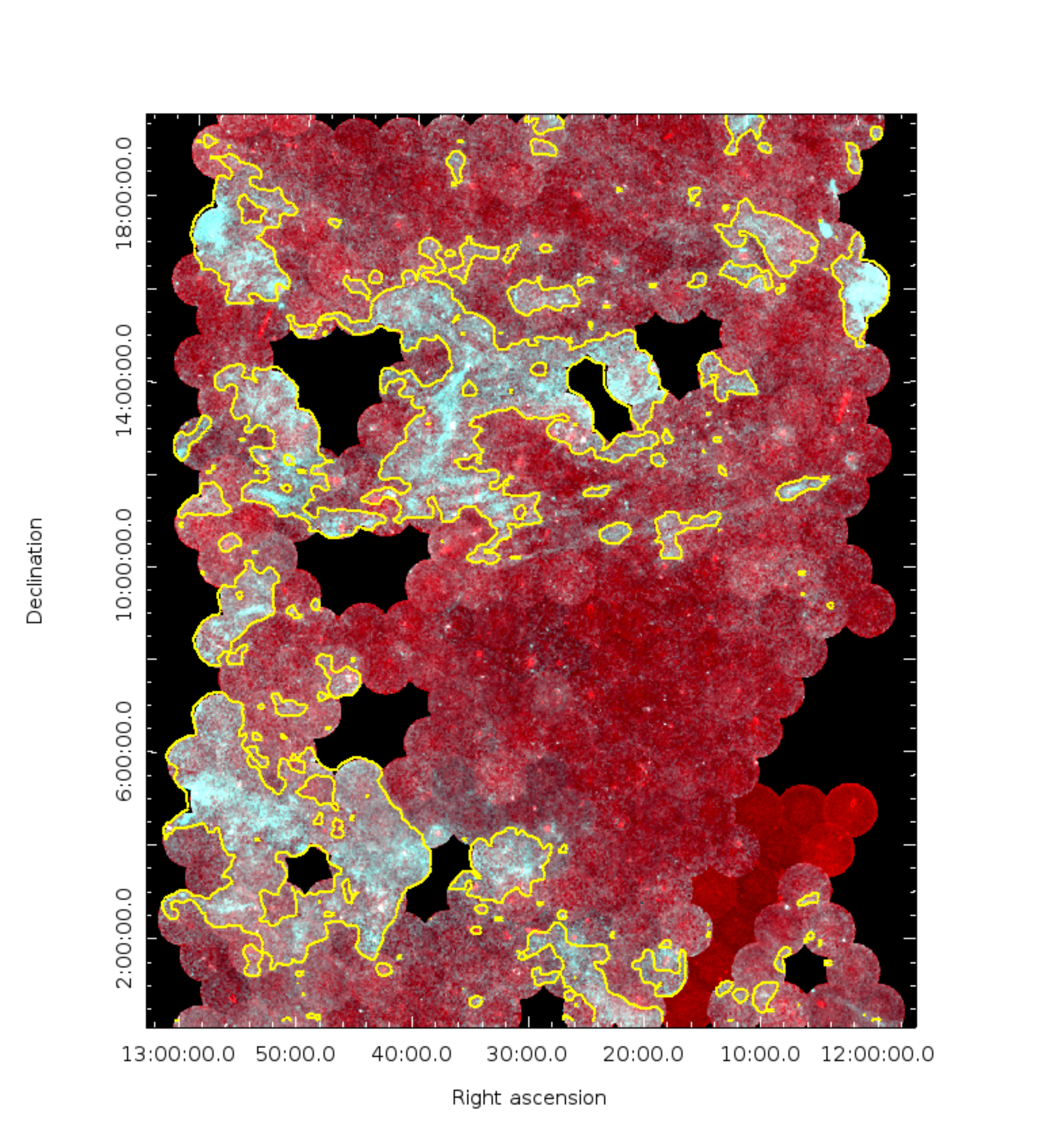} 
\includegraphics[height=9.cm]{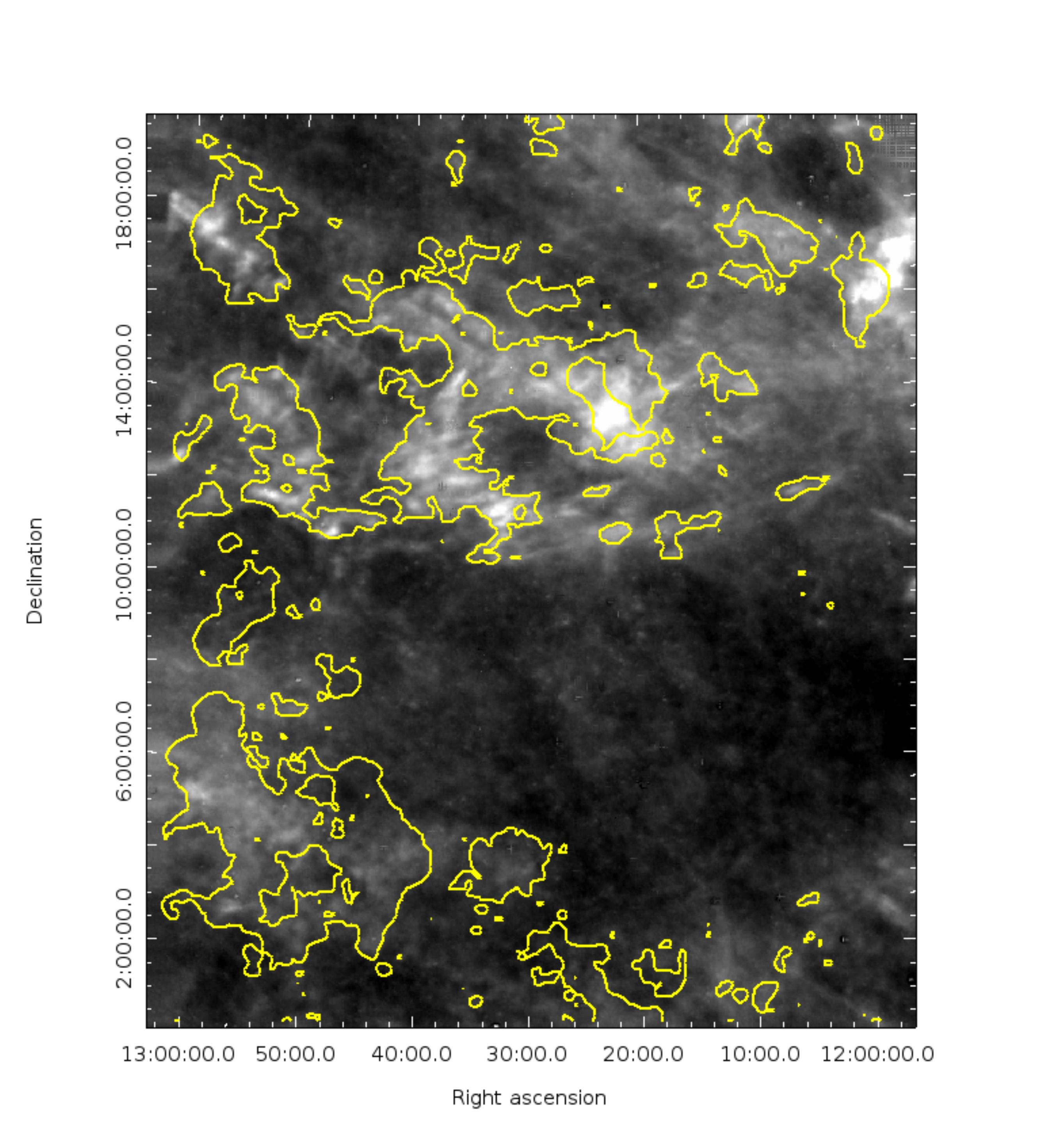} \\  
\includegraphics[height=9.cm]{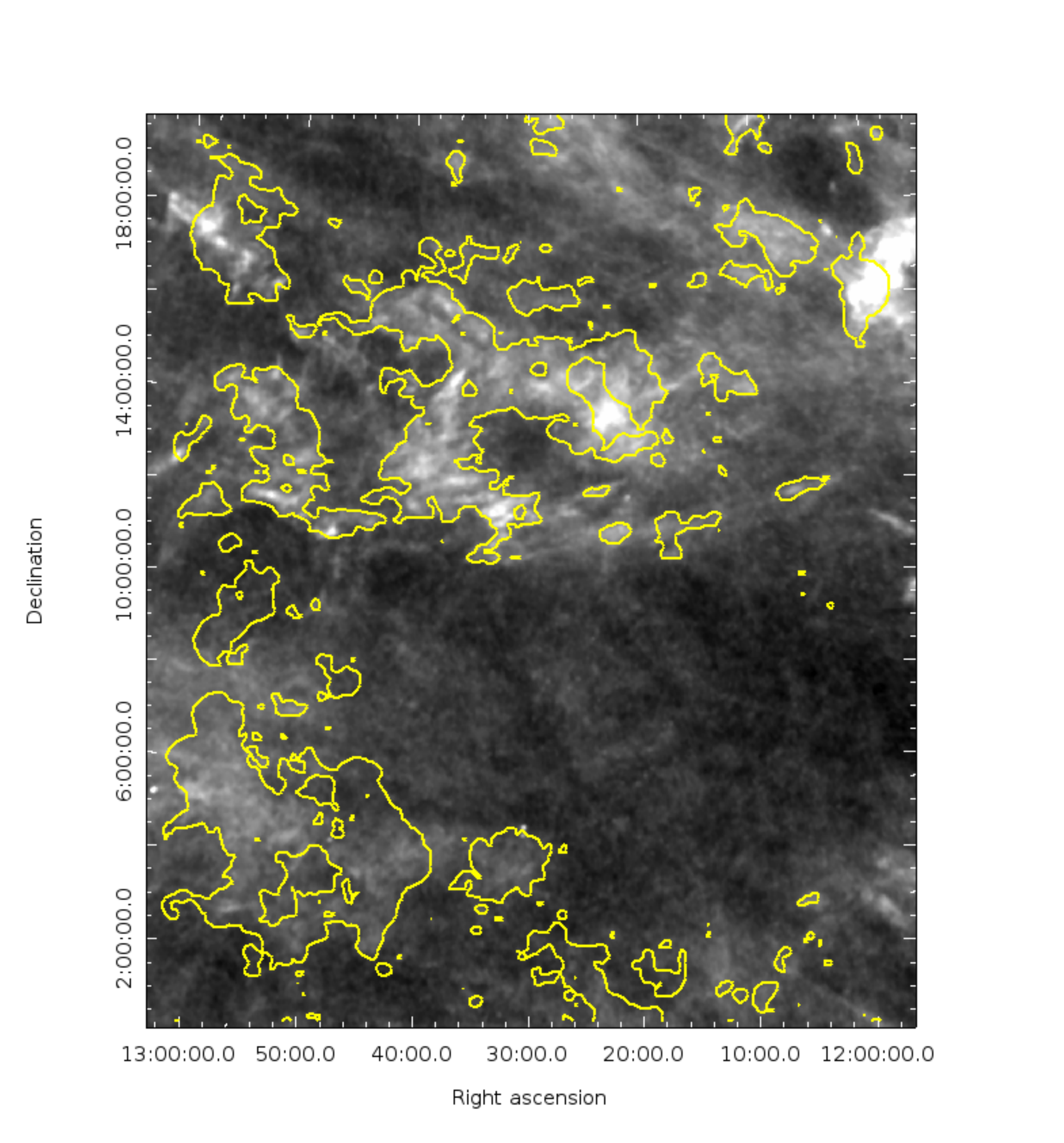}  
\includegraphics[height=9.cm]{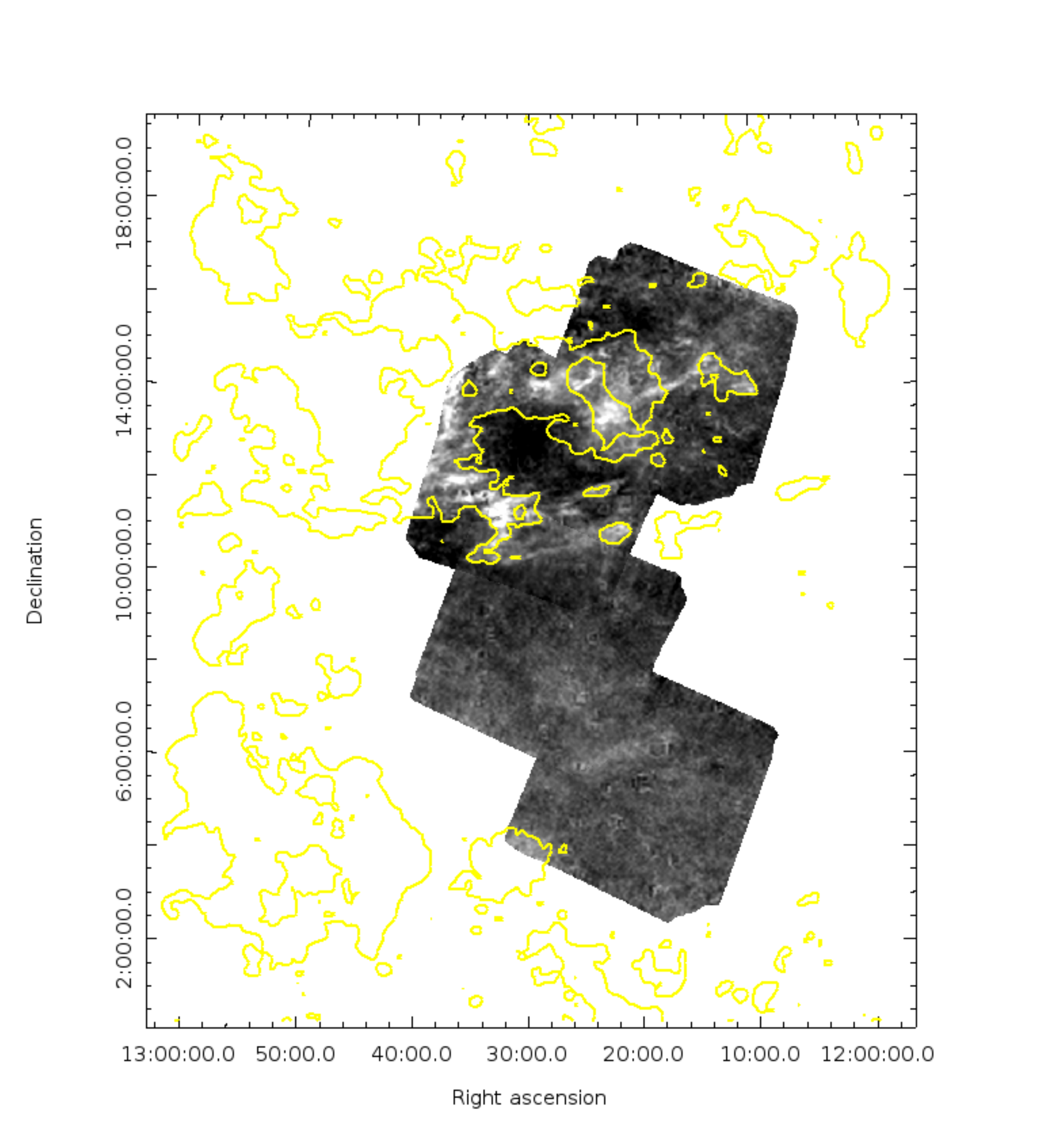} 
   \caption{Gallery of multi-wavelength tracers of the cirrus in the Virgo Area. Top-Left: composition of FUV (blue) and NUV (red) mosaics from GUViCS. Bottom-Left: \Planck{} \ebv{} map . Top-Right: 100 microns IRIS map. Bottom-Right: 250 microns HeViCS map. The yellow contours are the broadly defined regions of cirrus detected in FUV (Sect. \ref{sectioncontours}).}
     \label{FigAllwav} \label{FigFuvNuv}
    \end{figure*}

\subsection{Other wavelength maps}

Large scale variations in the sky background are known to be related to zodiacal light, and cirrus. 
These components are often studied by looking at their emission in the far infrared. 
\citet{schlegel98} produced maps of IRAS/COBE 100 $\mu$m emission that have been widely used in the literature.
The IRAS data were reprocessed by \citet{miville2005} to obtain at 100 microns a slightly improved angular resolution (4.3 arcmin), and a better calibration, taking into account the time and brightness response of the detector. This improved map is called IRIS (Improved Reprocessing of the IRAS Survey). \citet{planck2013}, however, mention that the zodiacal light subtraction was done differently in the IRIS project with respect to \citet{schlegel98}, and proposed a simple correction to use the large scale variations from \citet{schlegel98} and the small-scale variations from \citet{miville2005}. We use the IRIS 100 microns map and adopt the same correction in our study. 
\label{secirisdescrpt}
Finally, we prepared a mask with all VCC galaxies detected by IRAS and with some other bright objects such as known field galaxies and stars. The mask excludes up to twice the 100 microns IRAS resolution around each source. Masked regions were interpolated over.

The \Planck{} collaboration has used IRAS and \Planck{} all sky maps to compute the properties of the Galactic dust thermal emission. The collaboration made available the dust maps, including dust opacity, temperature and radiance \citep{planck11,planck2013}. The dust radiance was converted into a reddening \ebv{} map\footnote{\url{http://irsa.ipac.caltech.edu/data/Planck/release_1/all-sky-maps/}}  using the colours of SDSS quasars as a reference. This map  should be adequate to estimate the dust attenuation for extra-galactic sources, with a spatial resolution of about 5 arcmin. 

The HeViCS project \citep{davies2010} collected  \Herschel{} data in 4 fields within the area 
studied in this paper. In this work, we use the 250$\mu$m maps observed 
with the SPIRE instrument (spatial resolution of about 18 arcsecs). The data was reduced using the \Herschel{}  
dedicated software HIPE, adopting the v11 calibration tree, assuming 
the gain correction for extended sources, and combining the scans with 
the built-in destriper. This was found to preserve the large scale 
diffuse emission better than the dedicated pipeline used in other 
HeViCS works, which was tailored for extra-galactic targets (see, e.g. 
Auld et al. 2013). Further details on the data reduction, together 
with a study of the properties of the diffuse cirrus emission will 
be presented in Bianchi et al. (in preparation). 
We interpolated these maps over the same mask as the one used for the IRAS 100 microns emission.
However after this step, the HeViCS tiles are still filled with many faint sources. The background was thus determined following the same procedure as in \citet{ciro2014}, i.e. computing iteratively a background with SExtractor.
The cirrus are well visible in HeVICS images. \citet{baes14} discussed their role as potential non extra Galactic sources in their cross-correlation of HeViCS and \Planck{} maps.  A HeViCS study of the cirrus is in preparation (Bianchi et al.), thus the HeViCS findings are 
not discussed here in details but the data are shown for illustrative and comparison purpose. 
 
We extracted the parts of the 100 microns and \ebv{} \Planck{} map
covering our UV mosaics using the Aladin sky atlas \citep{bonnarell2000}. The 250 micron maps directly come from the HeViCS collaboration. 
All were registered to the same pixel scale as our various UV maps 
(1 arcmin and 20 arcsec).

In principle, it would be very interesting to compare the ultraviolet detection of cirrus to the all sky dust emission computed in the 
WISE W3 band by \citet{meisner2014}. At 12 microns, this map reveals the emission from small grains and Polycyclic Aromatic Hydrocarbon 
(PAHs) absorbing mostly radiation in the ultraviolet. Scattered ultraviolet light should then spatially occur close to the 12 microns emission
(while the thermal dust is heated by the general interstellar radiation field that can be spatially mismatched from the ultraviolet strong sources). We did download the dust emission maps published by  \citet{meisner2014}. Sadly, the Virgo area is crossed by a wide stripe of corrupted data, making them unusable for a statistical comparison that would correspond to our goals. A detailed comparison of some regions could still be possible in the future (we illustrate this point in section \ref{SecConclu}).

\section{Analysis of large scale variations}
\label{SecAna}

\subsection{Zodiacal light}

For general studies of the zodiacal light in the UV, the reader is invited 
to consult e.g. \citet{frey77,murthy14}. 
Here we only  analyse its effects on our mosaics and their contamination on the cirrus emission.
In the NUV band, we found that our object-subtracted map presented a strong gradient in its background, which can be ascribed to the zodiacal light. The \GALEX{} site proposes a zodiacal light tool\footnote{\url{http://sherpa.caltech.edu/gips/tools/zlct.html}} based on the zodiacal light study by \citet{leinert98}. It predicts the level of zodiacal light expected for a  position and an epoch. We checked with this tool that indeed a global decreasing gradient is expected in the NUV observations along the ecliptic latitude.

We show in Fig. \ref{FigBackvsLatNUV} the actual gradient with ecliptic latitude that we measured in our \GALEX{} 
diffuse NUV image over the full Virgo cluster area. It is compared to the prediction of the \GALEX{} zodiacal tool
for a few coordinates regularly covering the area, adopting the date of January 1st. Variations with the date do occur, but change in the same way at all latitudes. These variations are at most time lower than about 25 percent, except during a zodiacal emission spike, when GALEX observations were not executed. At low ecliptic latitude, we find large values of the background, consistent with the \GALEX{} tool model. At larger latitude, our background decreases, as the model does, but by a smaller amount. This may be due to either an under-estimation of
the zodiacal light model, or that other sources start to contribute to the background (ghosts of stars, artefacts,...). We note, however, that if the background was dominated by other sources, it should not depend on the latitude. The grey line in Fig. \ref{FigBackvsLatNUV} shows the gradient that we determined by hand to approximatively pass by the darkest regions within the map at different latitude. This gradient is only used in the next section to consider the cirrus NUV detection and in Fig. \ref{FigFuvNuv}.

We proceeded to the same exercise in FUV (Fig. \ref{FigBackvsLatFUV}). In this case, the expected level of the zodiacal light is extremely low. The observed level is much higher than the prediction and is independent of the latitude \citep[as found by][on a larger scale]{hamden2013}. Other effects thus dominate the background (e.g. ghosts, residuals from subtracted objects, unresolved distant sources, and Galactic cirrus studied in the following).

We did also look at how the diffuse FUV and NUV background counts vary with the \emph{Galactic} latitude. We found no 
trend with the Galactic latitude in FUV, but we investigate a limited range of Galactic latitude, above 55 deg, while most of the trend due to cirrus is expected to occur at lower latitudes \citep{hamden2013}. An anti-correlation is found in the NUV background counts, but we interpret it as 
the result of the variation with the ecliptic latitude (the zodiacal light described above), since in our field, the Galactic and ecliptic latitudes 
vary in the same direction.

\subsection{Ultraviolet cirrus detection}
\label{secdetection}

\label{secfuvbg}
In the FUV, we measured the background level in several regions among the 
darkest area of the sky in our mosaic. 
We did this by using the IRAF task IMEXAMINE in several rectangular regions, in a similar manner as previous works with \GALEX{} data \citep[e.g.][]{boissier07,boissier08,boissier12}.
As it can be guessed from Fig. \ref{FigBackvsLatFUV}, this background is found at a level of 0.25 counts per sec per arcmin$^2$ (with a 0.11 3-$\sigma$ uncertainty per pixel). We created 
a FUV backgroud-subtracted map by subtracting this constant value to the FUV mosaic.
By proceeding this way, we basically remove any emission not related to the cirrus themselves, 
that could be due to air-glow or extra-galactic sources, and that should correspond 
to the "offset" observed by \citet{hamden2013}. We acknowledge that this background could include
a very diffuse cirrus contribution that is then subtracted in our approach. The nature of this 
diffuse component emission is still unknown and the subject of several studies \citep[see e.g.][who test the 
hypothesis of a dark-matter interaction with the interstellar medium, or of small dust grains; and references therein]{henry14}.

The remaining FUV emission above this constant background is assumed to be due to cirrus.
It is true that some part of it could have other sources (extended halos of galaxies, tails or streams, intracluster light) 
but the result discussed in the rest of the paper seems to be consistent with this emission being dominated by the Galactic cirrus.
We created a detection map by adopting a 3-$\sigma$  threshold above this level. 
This map was then smoothed (by a Gaussian filter of 6 arcsec) with the only goal to 
compute contours (in ds9) at a single level (0.1 counts sec$^{-1}$ arcmin$^{-2}$) in order to geometrically identify broadly the cirrus regions. These contours are not directly used in the science analysis 
of this paper, but we overlay them in our figures. 
They are shown in Fig. \ref{FigFuvMaps} over the the FUV mosaics obtained from AIS data and from deeper GUViCS data.

\label{sectioncontours}

\label{secdefinenuvbg} 
In the NUV, we first constructed a background map to reproduce the gradient observed in Fig. \ref{FigBackvsLatNUV}. We used the darkest regions to estimate its zero-point.
This background was subtracted to the NUV image. We then inspected the sky level within this background-subtracted image, and found it to be flat.  Measuring the sky level in dark regions we obtained the value of 0.1 counts per sec per arcmin$^2$ with a 3-$\sigma$ value of 0.52: it is consistent with zero, demonstrating that the sky gradient subtraction procedure performed well.
We then produced a NUV "detection map", by adopting also here a threshold at  this 3-$\sigma$ value.

\subsection{UV Properties of Galactic cirrus}

\begin{figure}
   \centering  
   \includegraphics[width=10.5cm]{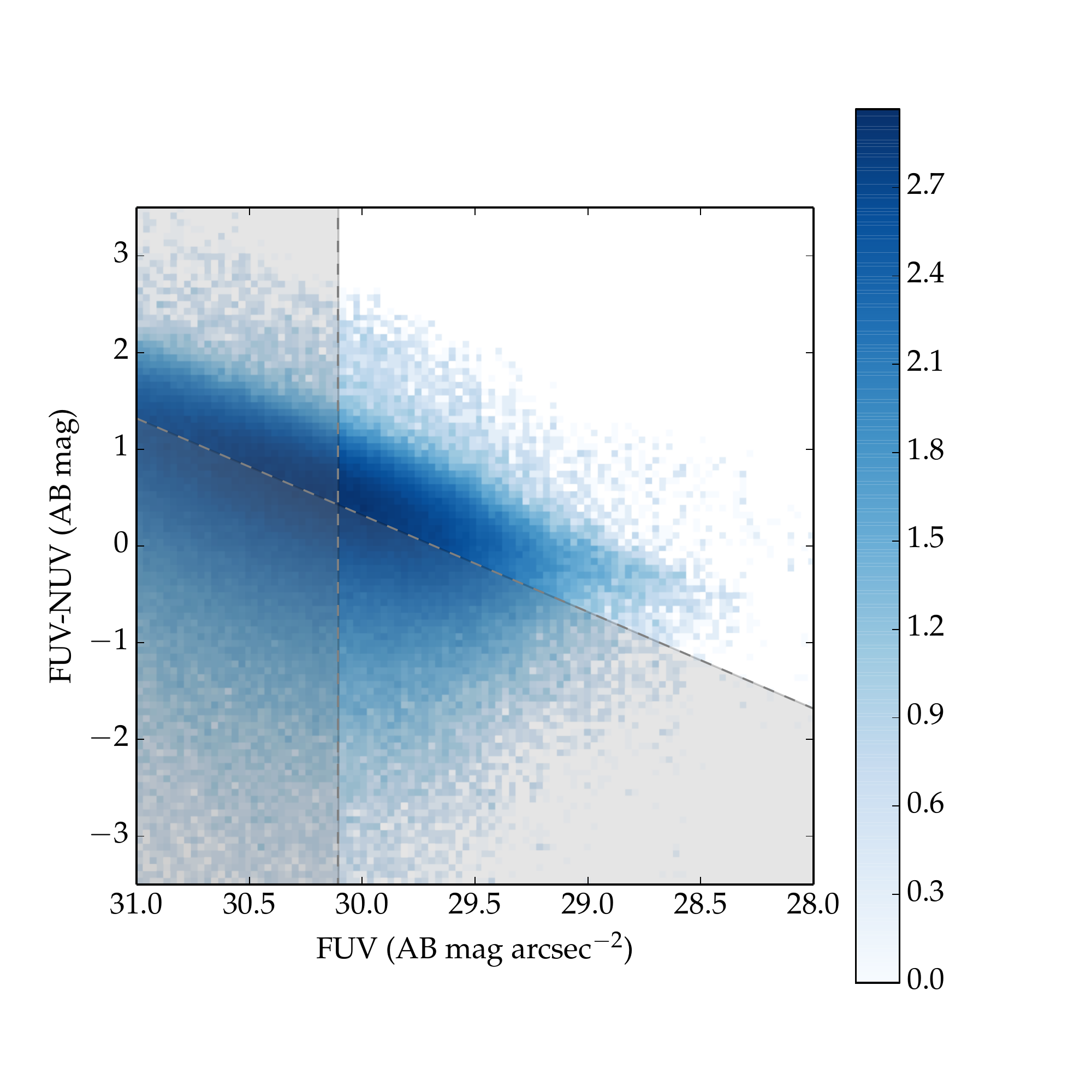} 
   \caption{FUV-NUV colour as a function of surface brightness (based on the 1 arcmin pixel size mosaics). 
The lines shows our detection thresholds in FUV and NUV (shaded area are below the detection threshold and thus subject to large uncertainties). The colour-bar indicates the decimal logarithm 
of the number of pixels in each 2D bin.}
     \label{FigFuvNuvHisto2D}%
    \end{figure}

Fig \ref{FigFuvNuv} (top-left) shows the backgroud-subtracted images in NUV and FUV with the FUV detection contours. The comparison definitively shows the blue colour of the cirrus with respect to the background.
To characterize the UV properties of our cirrus, we show in Fig.   \ref{FigFuvNuvHisto2D} the 
relation between the FUV-NUV colour and the FUV surface brightness, obtained from the object-subtracted, backgroud-subtracted 3x3 median-filtered 1 arcmin size pixel maps. The higher resolution maps (having a higher signal to noise, but smaller and inhomogeneous coverage) produce very similar results from this point of view.
Our data point to bluer colours for brighter surface brightnesses. \citet{sujatha10} found such an increase in the FUV/NUV ratio that they interpreted as a sign of fluorescent emission of molecular hydrogen. In our data, however, we cannot rule out a detection issue, as can be seen from Fig. \ref{FigFuvNuvHisto2D}. 
Taking into account the thresholds discussed in Sect. \ref{secdetection} and restricting 
ourselves to the pixels where the signal to noise is larger than 3 at both wavelengths,
we find that the typical FUV-NUV colour of the cirrus is close to 0. This is very similar to the colour of FUV-NUV $\sim$ 0.11 that can be deduced from the spectral distribution found by \citet{gondhalekar80} for the 
interstellar radiation field, probably typical of B-stars illuminating the dust clouds. In the scattered haloes of external star-forming galaxies, \citet{hodges2014} also found quite blue colours with SEDs ($\lambda F_{\lambda}$) that are flat (or declining from FUV to NUV), corresponding to a FUV-NUV colour of $\sim$ 0.4 (or bluer).

\begin{figure*}
   \centering  
\includegraphics[width=18cm]{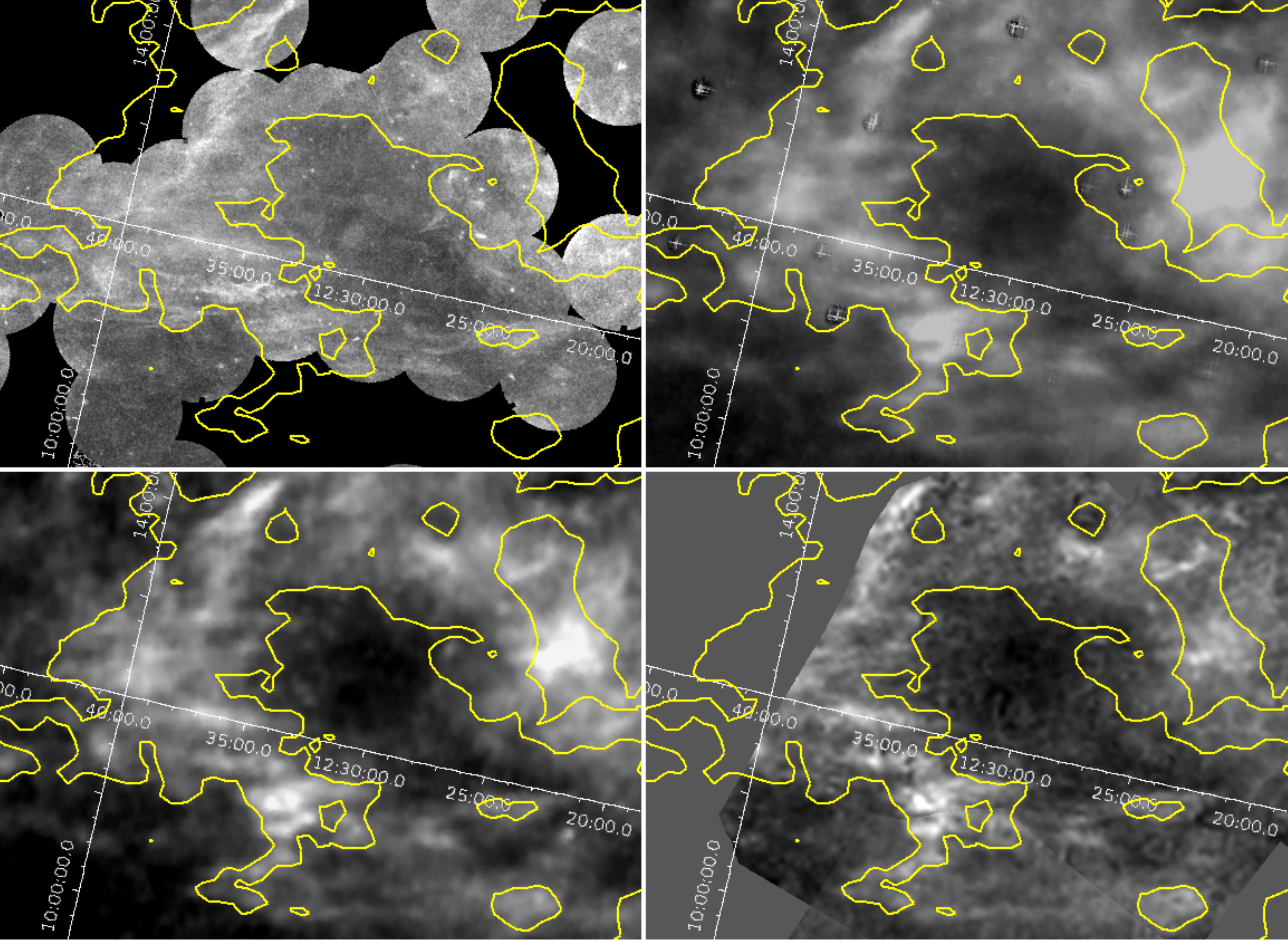}
   \caption{Comparison of several cirrus indicators in a zoomed region with deep FUV data (top-left). This map is compared with the 100 microns IRIS map (top-right), HeViCS 250 microns (bottom-right), \Planck{} \ebv{} map (bottom-left). }
     \label{Figcompa}%
    \end{figure*}

\subsection{Comparison to other wavelengths}

Fig. \ref{FigAllwav} shows the 100 microns, 250 microns, and \Planck{} dust reddening maps 
compared to a FUV+NUV combination (obtained from the shallow data covering a large fraction of the full area). 
Globally, a similar morphology is found with the various tracers. 
Such an agreement between various indicators for the cirrus presence has been already discussed before e.g. by 
\citet{guha94} for the optical, 100 microns and gas distribution, or \citet{seon11} for relations between FUV, gas column density, 100 microns radiation. 
The agreement between the various tracers is, however, not perfect. 
For instance, some regions of the \Planck{} \ebv{} or far infrared maps are dark whithin UV contours in Fig. \ref{FigAllwav} (e.g. around 12h55m, 9 degrees).
Small-scale variations do bring clues
about the physical properties of the dust \citep{guha94} even if another source of dispersion is due to 
the anisotropic interstellar radiation field and the precise geometry of dust and bright stars \citep{witt97,seon11}.
It is beyond the scope of this paper to explore these aspects in details here,
but the maps that we distribute may help to explore such studies in the future.

We show in Fig. \ref{Figcompa} a comparison of several dust tracers in a region for which we do have data deeper than 
the AIS, using our 20 arcsec pixel map. It reveals that the UV allows to follow in details the structure of the cirrus
with a better resolution than the traditional tracers.

\begin{figure}
   \centering  
   \includegraphics[width=10.5cm]{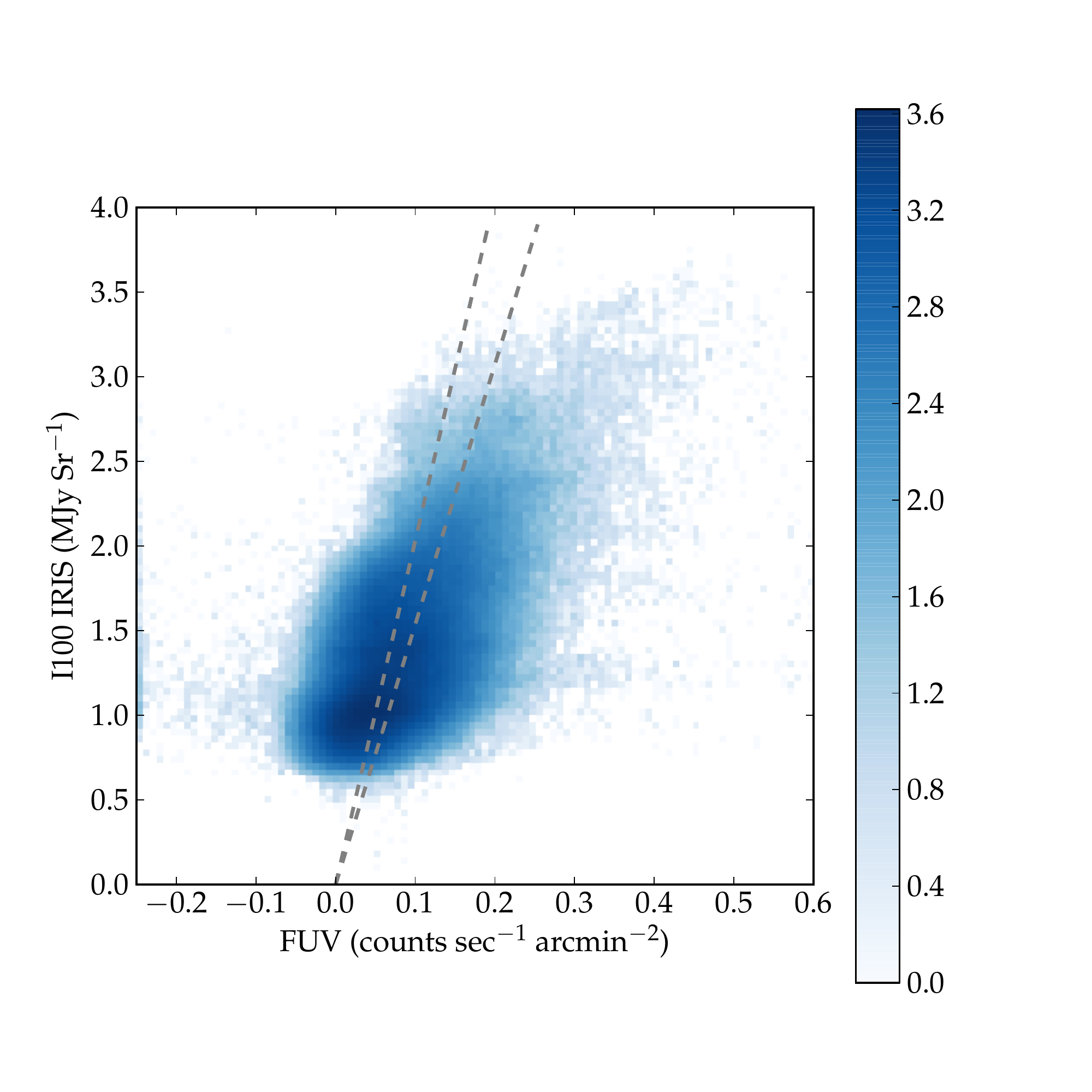}
   \caption{FUV counts vs 100 microns intensity diagram (I100).
The colour scale shows the logarithm of the numbers of pixels within 2D-bins. The grey lines indicate the slope found by \citet{hamden2013} for two Galactic latitudes corresponding to the GUViCS area. The maps used have 1 arcmin size pixels and have been median-filtered over 3x3 pixels. The 100 microns correspond to IRIS reprocessed IRAS data corrected on large scales as explained in section \ref{secirisdescrpt}.}
    \label{Fig100microns}%
    \end{figure}

\begin{figure}
   \centering  
   \includegraphics[width=10.5cm]{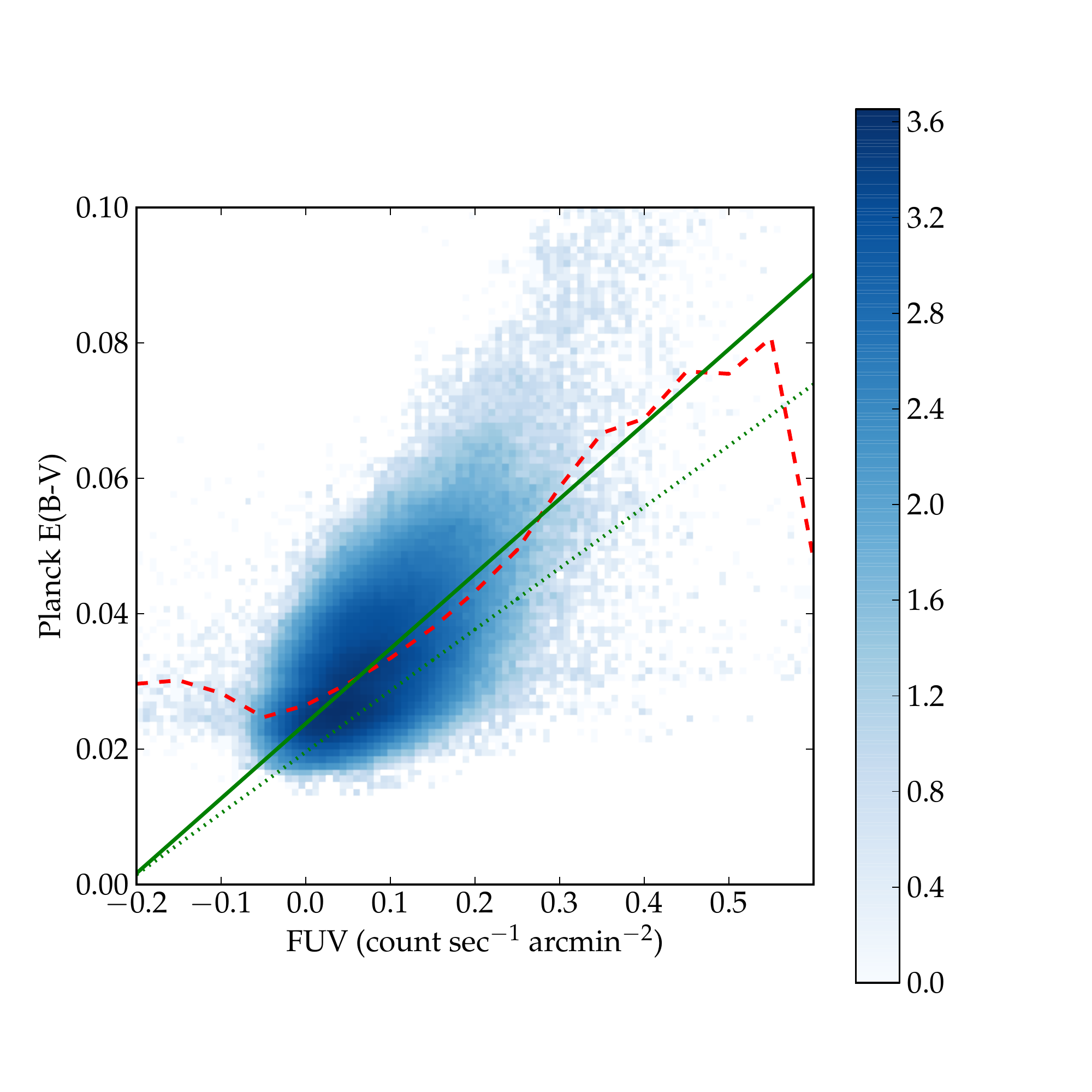} 
   \caption{\Planck{} \ebv{} vs FUV counts. The colour bar indicates the logarithm of the number of pixels within 
each 2D-bins. The dashed line shows averages within FUV flux bins. The solid line indicates a simple linear fit to the average trend. The dotted line indicates the result of a similar analysis based on the \citet{schlegel98} \ebv{} map. }
 \label{FigEBV}
    \label{FigPlanck}%
    \end{figure}

Beyond a simple visual comparison of the maps, we now proceed to a quantitative comparison, pixel by pixel, of the FUV cirrus emission and other dust tracers.  Fig. \ref{Fig100microns} and \ref{FigPlanck} show the distributions of pixels, 
and Table 1 provides the correlation coefficients between the FUV and other dust tracers pixels.

A relation between FUV and 100 microns emission had been reported in several studies 
\citep{perault91,witt97,sasseen96}. We show the relation we find in the Virgo direction in Fig.  \ref
{Fig100microns}. The relation found by \citet{hamden2013} on larger scales seems to holds at our 1 arcmin scale, with a 3x3 pixels median-filter (an adequate pixel size taking into account the 100 microns resolution). In this comparison, we used only the slope of the 100 microns-FUV relation of \citet{hamden2013} since the intersect is believed to be due to uniform sources that we subtracted in our map construction. We adopt their slopes in the Galactic latitude ranges corresponding to Virgo (60-75 and 75-90). 
The good agreement suggests that our background measurement and subtraction was efficient. On the other hand, we note an important dispersion, a sign that the dust emission in the far infrared and the FUV reflection on dust are non perfectly coincident.
We found a similar relation between the FUV and the 250 microns emission, albeit with more dispersion. The 250 map, however, covers only a part of our FUV data.

\label{sectcomplementarity}

\begin{figure}
   \centering  
   \includegraphics[width=8.5cm]{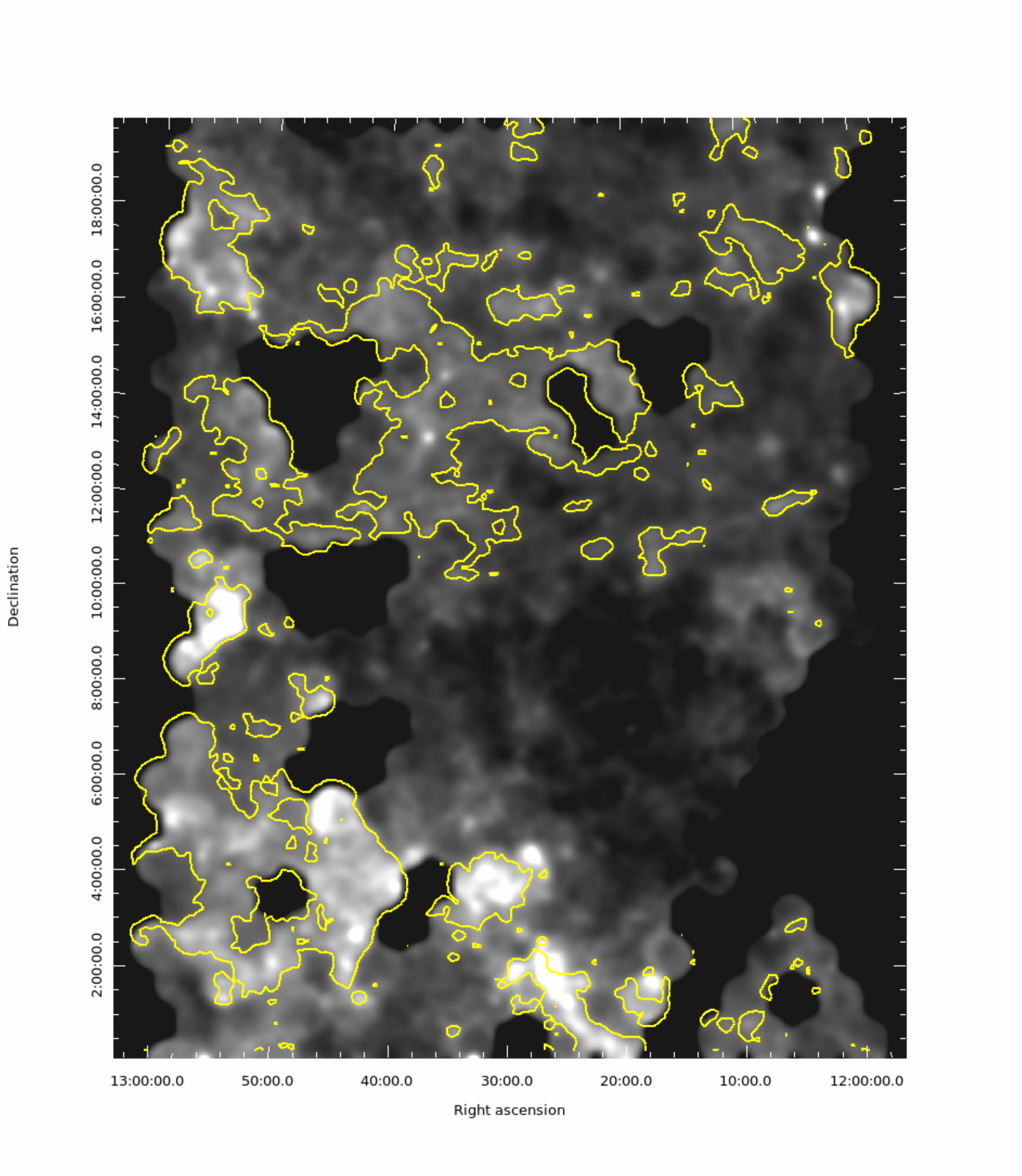}
   \caption{Ratio of the FUV diffuse emission (in the 1 arcmin pixel map) to the 100 microns emission, after applying to both image a Gaussian filter of 6 arcmin. The contours indicate roughly the regions in which the cirrus emission is found (Sect. \ref{sectioncontours}). }
    \label{FigComplementarity}%
    \end{figure}

\begin{figure}
   \centering  
  \includegraphics[width=8.5cm]{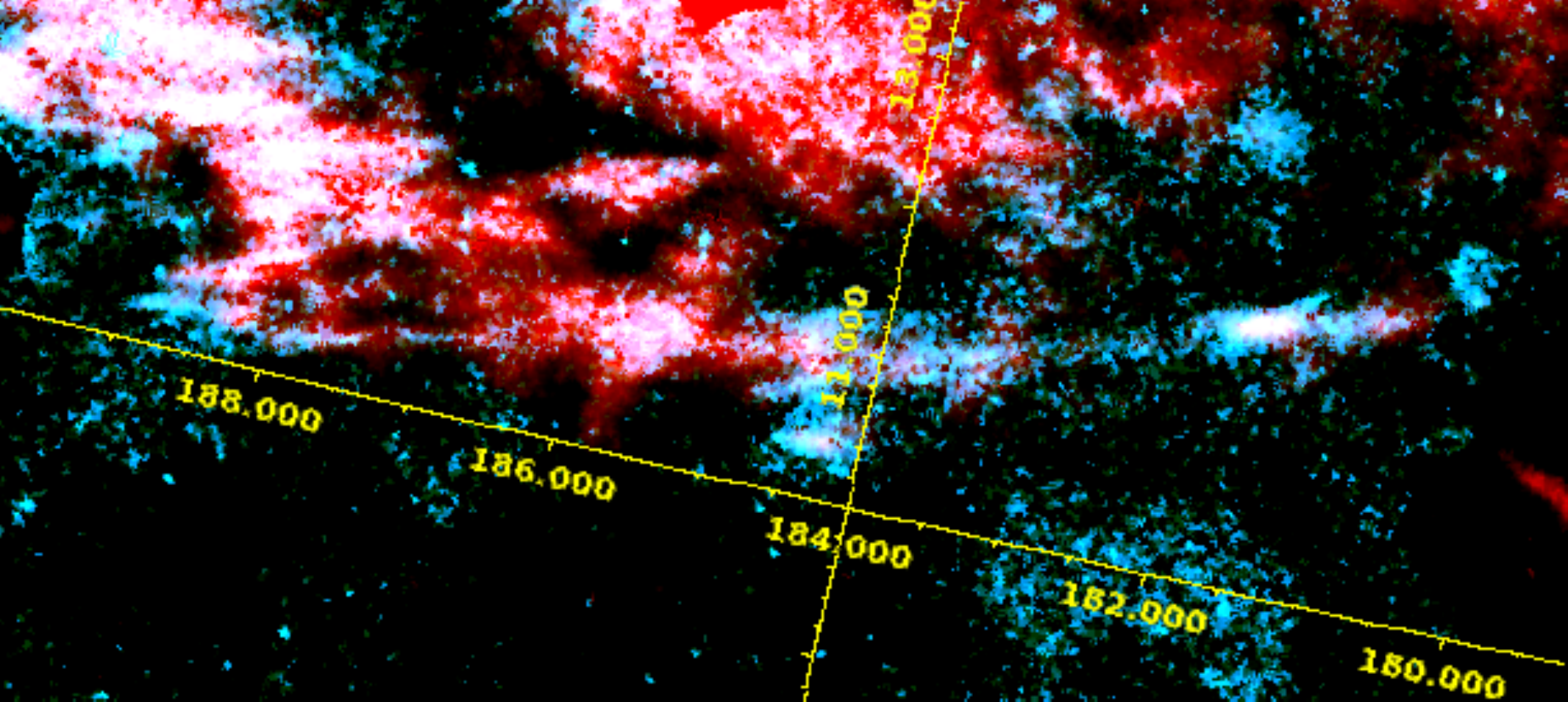}
   \caption{Combination the FUV emission (blue) and 100 microns emission (red) along an extended cirrus feature, spreading from left  to the right part of the figure. The scaling was chosen to show the complementarity of the two dust tracers, since the colour is changing along the structure.}
    \label{FigComplementarity2}%
    \end{figure}

The dispersion in the relation between the FUV and infrared pixel values (reflected by the relatively poor correlation coefficients) is likely due to the fact that while the two tracers globally trace the shape of the cirrus, they are more complementary than equivalent.
This can be better seen in Fig. \ref{FigComplementarity} in which we show the ratio of the FUV diffuse emission to the IRIS 100 microns emission. We first applied a 6 arcmin Gaussian convolution to each image to reveal the differences on large scales and avoid being dominated by small scale fluctuations (or resolution effects). It is evident from this figure that this ratio is not identical in the field but varies with position. It is not only due to noise since it varies also within the contours indicating regions where FUV is definitively detected.
Fig. \ref{FigComplementarity2} shows a zoom in a region where a clear elongated cirrus structure is seen, part of it being obvious through its FUV emission, part of it through its far infrared emission. 
Some of the mismatch could be due to the relative geometry of stars and various type of dust. The FUV scattered light may occur on small grains predominantly in regions close to FUV emitting stars, while the far infrared radiation corresponds to the thermal emission of the dust heated by the interstellar radiation field, heating that can also occur in the absence of the hottest stars in the immediate vicinity.

\begin{table}
\caption{Pearson correlation coefficients}
\centering
\begin{tabular}{l l l }
\hline
\hline
X  & Y  & Correlation coefficient \\
variable & variable & (Pearson) \\
\hline
FUV & 100 microns   & 0.57 \\
FUV & 250 microns & 0.47 \\
FUV & \ebv{} \Planck{} & 0.61 \\
\hline
\end{tabular}
\end{table}

\subsection{FUV as a dust tracer in the Virgo Cluster area}
\label{secdustfuv}

We show the distribution of pixels in the  FUV emission vs \citet{planck2013} \ebv{} extinction plane in Fig. \ref{FigPlanck}. 
The two variables are correlated. A very similar trend is obtained on the basis of the \citet{schlegel98} maps 
(the figure includes the best linear fit in this case).
This indicates that the FUV diffuse emission is at least loosely linked to the extinction as measured through the dust emission. We explore quantitatively the link between the FUV diffuse emission and the extinction in this subsection.
The FUV map used is the 1 arcmin pixel one, median-smoothed over 3x3 pixels. The \Planck{} extinction map has a resolution of a few arcmin, thus this scale is pertinent for our comparison (we however computed the distribution at higher resolution -with the 20 arcsec pixels map- and found very similar trends).
As expected, since the FUV and the \Planck{} maps do not correspond perfectly to each other, some scatter is found.
The figure, however, suggests a clear average relation between the two quantities. We computed averages within bins of FUV emission  and performed a simple fit to these values to find:
\begin{equation}
\label{eqfuvext}
E(B-V)= 0.02378 + 0.1105 \times FUV {\rm (counts \, s^{-1} \, arcmin^{-2})}
\end{equation}

A shallower slope ($\sim$ 0.09 instead of 0.11) is found if the \citet{schlegel98} map is used rather than the \citet{planck2013} one. Considering the dispersion, the value of this slope is uncertain, but a correlation is clearly present.
Assuming the underlying average relation  between FUV emission and dust extinction  holds at small scales, we invite researcher working in the Virgo area to use our distributed 20 arcsec (or 1arcmin) map and this relation to predict "alternative" extinction values with respect to the traditional \citet{schlegel98} ones (or \Planck{} ones). The advantage is that these FUV maps reveal detailed structures of the cirrus not visible in the long-wavelength maps (lack of resolution/sensitivity, and complementarity : see Sect. \ref{sectcomplementarity}). 
We warn the reader however: a 1 to 1 relation is not expected as the the FUV emission is sensitive on the geometry of dust and UV-emitting stars
\citep{witt97,seon11}, 
Equation \ref{eqfuvext} should then be used with caution, especially for individual objects.
Despite this drawback, we believe that an attenuation computed from our FUV diffuse maps provides an alternative value of the reddening that may be complementary to traditional dust tracers. 

\subsection{Testing the FUV extinction determination with an extra-galactic catalogue}

\label{sectestebv}

To test this new extinction estimation, we use background galaxies detected in NUV using the GUViCS
catalogue from \citet{voyer2014}. The idea to use background objects to study Galactic extinction is not new
\citep{burstein78} and was applied by \citet{fukugita2004} to SDSS data. 
In fact, the \ebv{} \citet{planck2013} map is based on the same idea applied to the colours of background quasars used to calibrate the relation between \ebv{} and the thermal dust emission.
\citet{cuillandre01} applied a similar method to measure attenuation in M31 from background galaxies. They also explored the effect of attenuation on their colour and found it to be hard to detect. In our case, however, 
the used colour is more favourable as we expect a larger effect, using the NUV band, very sensitive to extinction.

\begin{figure}
   \centering  
   \includegraphics[width=8.5cm]{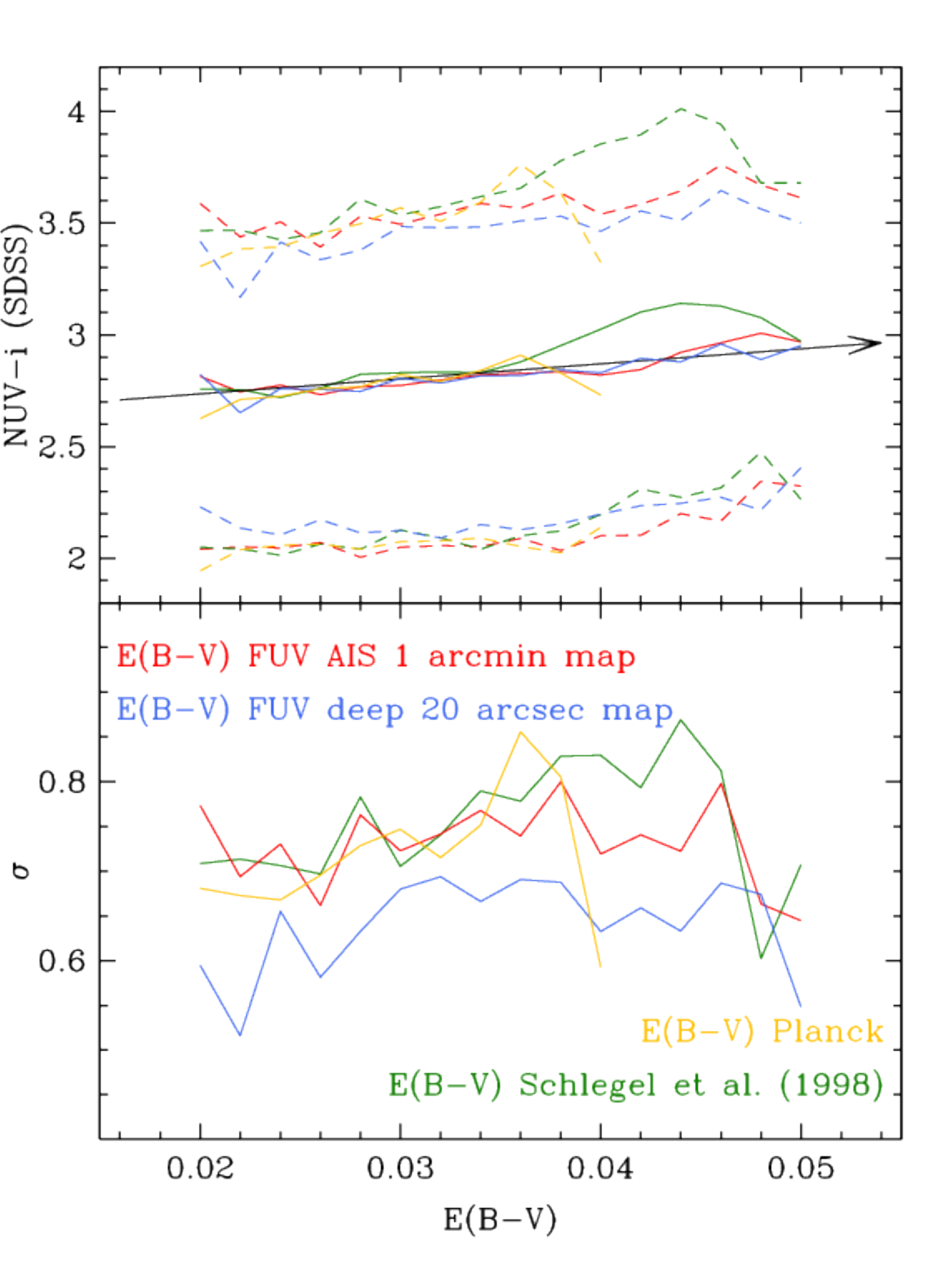}
   \caption{Top: NUV-i colour of SDSS galaxies in the \citet{voyer2014} catalogue as a function of the Galactic dust reddening (taking only objects with signal to noise larger than 10). The solid lines show the average, while the dashed lines indicate the 1-sigma dispersion. The dust reddening is derived either from the Schlegel maps (as given in the GUViCS catalogue), from the Planck \ebv{} map, or derived from our FUV 1 arcmin or 20 arcsec resolution diffuse emission maps. The arrow indicates the expected reddening trend (see Sect. \ref{sectestebv}). Bottom: the 1 sigma dispersion of the NUV-i colour is indicated as a function of the reddening. The range plotted in both panel corresponds to the bins where we do obtain \ebv{} values for each method.}
    \label{FigCatalogColors}%
    \end{figure}

We do not determine the extinction from the galaxy counts, but rather verify that the colours 
of background galaxies are statistically in agreement 
with the Galaxy dust reddening at their position.
In Fig. \ref{FigCatalogColors}, we show the trend between the $NUV-i$ colour and reddening for Virgo and background galaxies, with a signal to noise larger than 10 in the Voyer et al. catalogue, leaving us with around 25000 galaxies (similar results are obtained when the minimal signal to noise is reduced to 2, with about 70000 objects). This number is of course much larger than the number of quasars in the same region. The galaxy colour is directly taken from the catalogue. The figure shows the average and the dispersion of the colour of galaxies in various bins of Galactic reddening. We show four cases depending on which reddening is used: the one from the \citet{schlegel98} maps, given in the catalogue; the one from the \emph{Planck} \ebv{} map;  or the one derived from the FUV maps (the 1 arcmin and 20 arcsec pixel ones) applying equation \ref{eqfuvext}. In all cases, the average colour becomes redder (with a broad dispersion) with the reddening. At least statistically, using the \ebv{} derived from the FUV produces the expected trend on the colour, based on a Milky Way extinction curve \citep{cardelli89} with :  $A(NUV)=8.74 E(B-V)$ and  $A(i) = 1.98 E(B-V)$.
It is hard to distinguish the relative performances of the various reddening 
maps since the dispersion in the colour of galaxies (linked to their star 
formation histories, metallicities, ages,redshift) is much larger than the reddening trend itself.
The observed variation of the $NUV-i$ colour with \ebv{} is consistent with the one predicted 
simply with this Milky Way extinction law, suggesting that the various extinction tracers provide a similar level of accuracy.

Looking into the details, however, it can be seen in Fig.  \ref{FigCatalogColors} that the dispersion of the colour at a given \ebv{} 
depends slightly on the source of dust reddening considered. The one derived from the FUV 20 arcsec map is smaller than the other ones by about 10 percent. The dispersion of the $NUV-i$ colour of background galaxies in this diagram has two components, one is the intrinsic spread of colours of galaxies (due to their various star formation histories, metallicity, redshift) that is the same in each cases. The other one is obviously the incorrect assignment of the Galactic attenuation, which is changing depending on the dust tracer chosen. The ``best'' \ebv{} estimator should produce the smallest dispersion in this Figure (for a perfect estimator, we would be left only with the dispersion intrinsic to galaxies). We can thus conclude that the \ebv{} derived from the FUV 20 arcsec map does statistically a better job for our sample than the other ones, although the difference is small. We verified that this result is not due to the smaller area covered by this mosaic: repeating the exercise on the 1 arcmin map restricted to the area used in the  20 arcsec study does not reduce its dispersion. We thus have to attribute this better result to the resolution and depth gain when using the 20 arcsec pixel FUV map to estimate \ebv{}.

Some caution should be used when using this attenuation estimator. While we found that the diffuse FUV light correlates with E(B-V), and we use this fact to "calibrate" the redenning, it is clear that the FUV scattered light does not provide information on small grain absorption in FUV. Moreover, recent works suggest a large variability in the dust properties at high Galactic latitudes \citep[][]{ysard2015}.
\citet{peek2013} hint in addition to deviations from the standard extinction curve in the FUV band. Deriving extinctions in the FUV band from our estimate may thus produce erroneous results.

   \begin{figure*}
   \centering  
  \includegraphics[width=4.5cm]{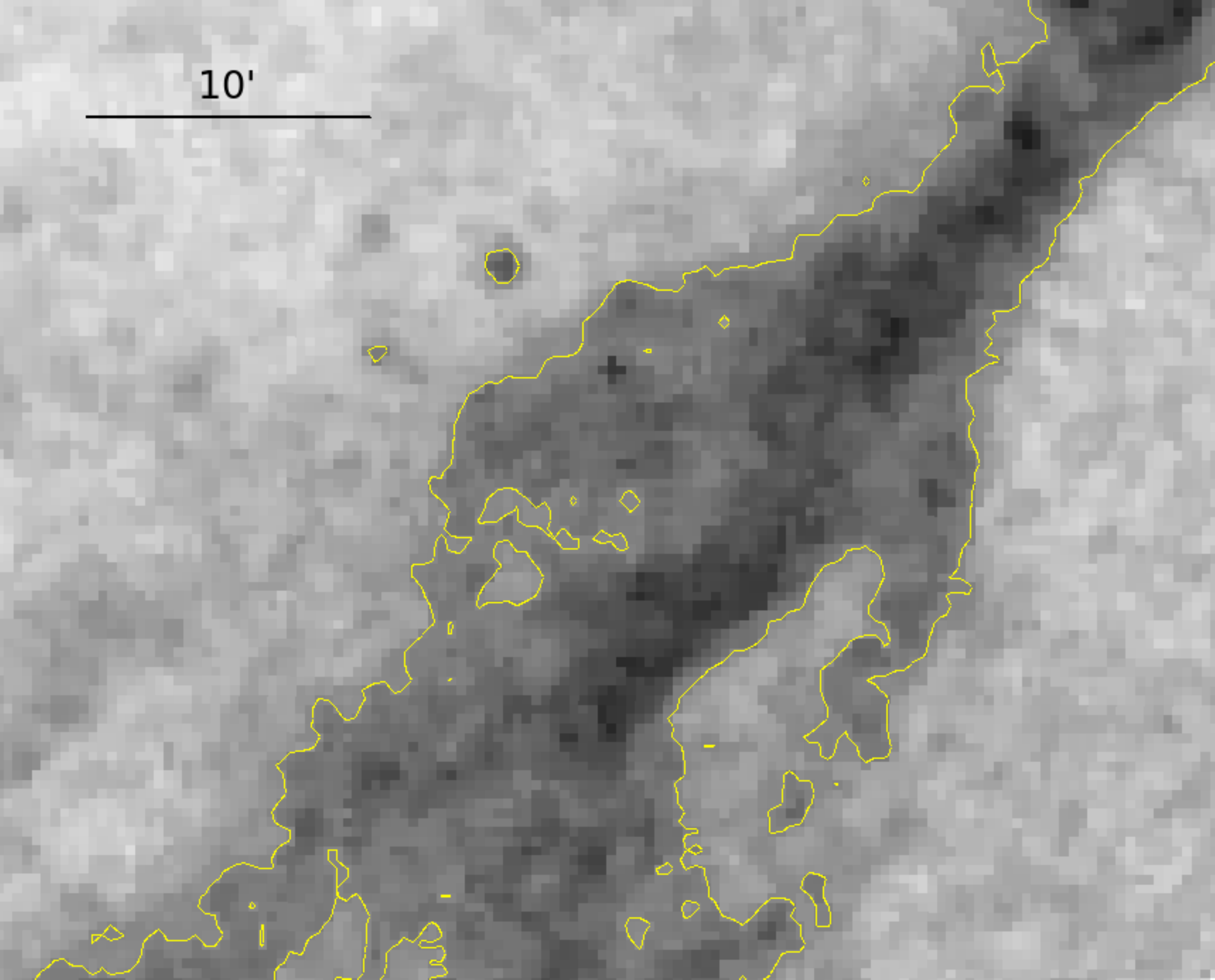} 
  \includegraphics[width=4.5cm]{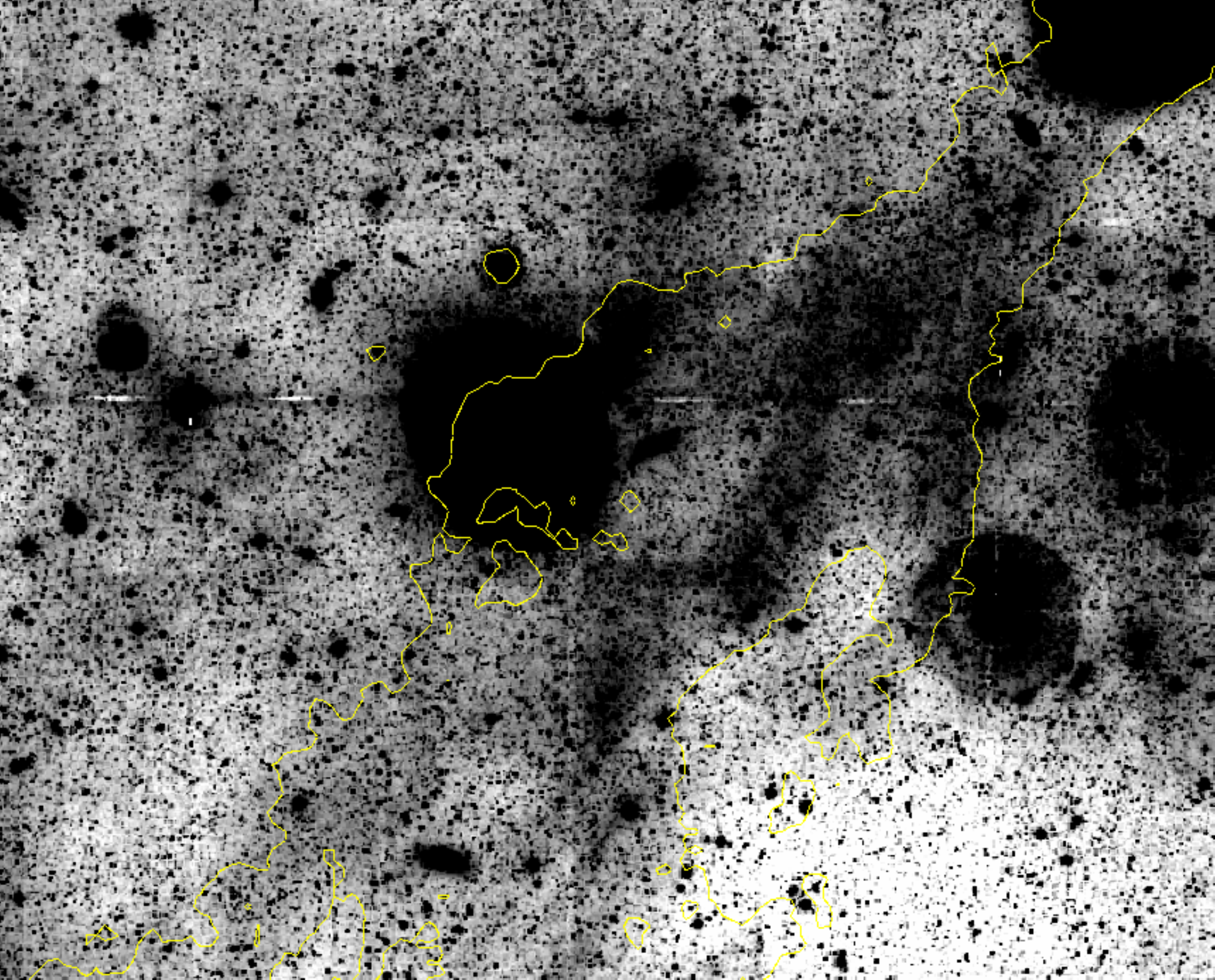} 
  \includegraphics[width=4.5cm]{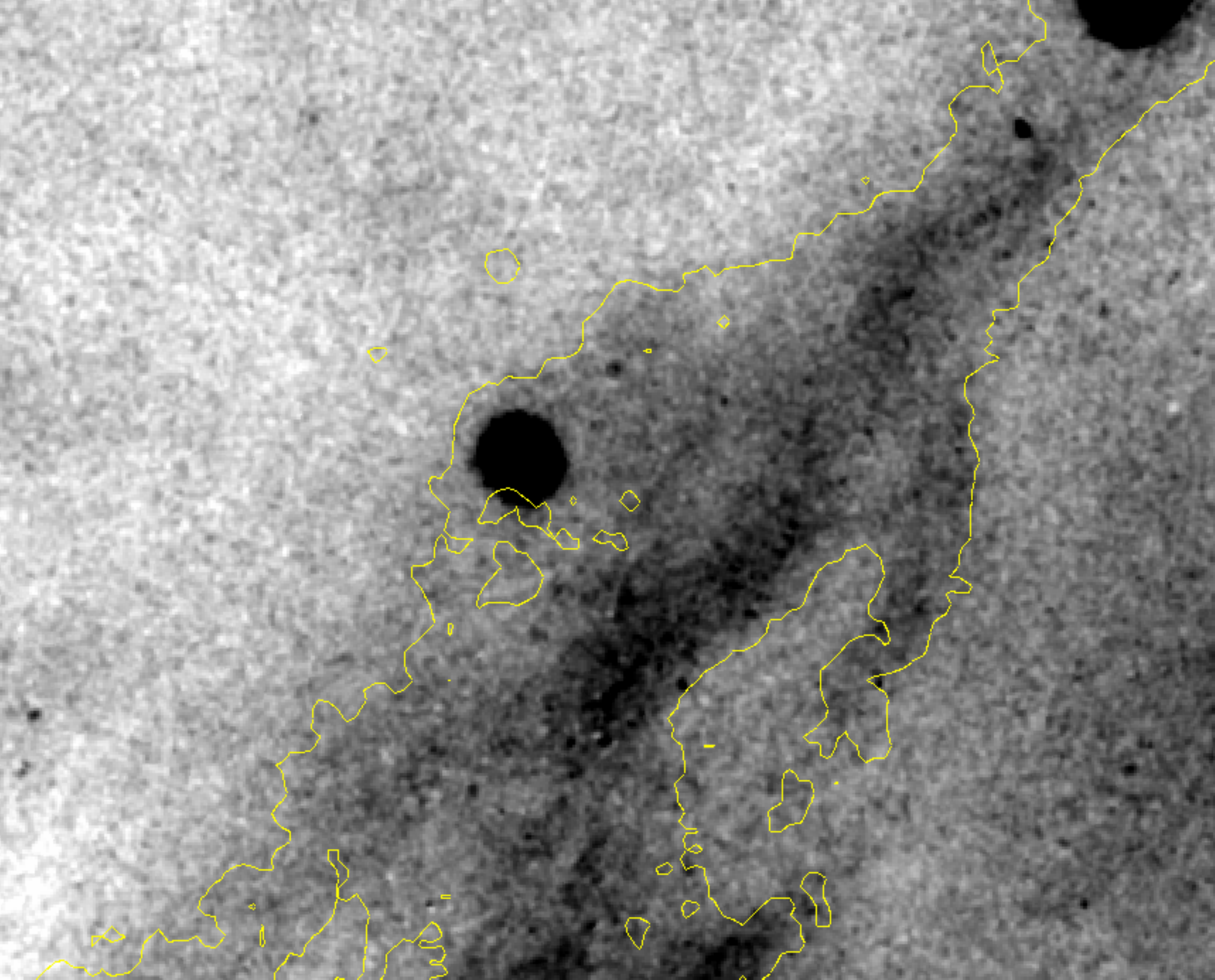} 
   \caption{A region around coordinates RA=12:36:07.201, DEC=+14:16:39.98 is shown as seen  in the 20 arcsec pixels FUV diffuse light map (left),
 in the NGVS \citep{ferrarese2012} $g$ band image (middle), and in the WISE W3 diffuse emission (right)  from \citet{meisner2014} on the same scale. To ease the comparison, a single contour was computed in FUV and overlaid on the other images.}
    \label{FigNgvsFuv}%
    \end{figure*}

\section{Conclusion}
\label{SecConclu}

We characterised the diffuse FUV emission in the Virgo cluster direction with the GUViCS data. We compared this emission (geometrically and statistically) to the NUV emission, IRIS 100 microns emission, HeViCS 250 microns emission, \ebv{} from \citet{schlegel98}
and from the \Planck{} collaboration.

Cirrus are clearly detected in the FUV channel. They are characterised by a constant blue FUV-NUV $\sim$ 0 colour
(an observed trend with FUV surface brightness is in fact related to observational limits).

We find an overall agreement between the locations of cirrus seen in FUV with those from other tracers, although some distinctions are found; certain regions are visible in the FUV maps but do not appear obviously in the infra-red or \Planck{} maps and vice versa. Despite these regions, a correlation is found when the various cirrus indicators are compared on a pixel basis (with, however, a large dispersion).
The FUV reveals details in the structure of the cirrus that are not always seen at any other wavelength,
owing to a good spatial resolution and sensitivity, but also to a complementary between the FUV and far infrared emission.

The diffuse FUV emission maps produced in this work may be useful for many purpose:

\begin{itemize}
\item  The study of the dust properties in the UV (albedo, scattering), or the properties of cirrus in general with multi-wavelength maps matched at the same pixel size. For instance, a small region with both  deep FUV data and good quality WISE W3 dust emission is shown in Fig. \ref{FigNgvsFuv}. The good agreement between the two tracers could confirm that the 12 microns emission comes from small grains and PAHs, which absorb the FUV radiation.

\item In conjunction with deep optical data, the study of the Extended Red Emission phenomenon.

\item Identification of regions "polluted" by faint structures related to Galactic cirrus. This could be very useful for the exploitation of the surveys coming to an achievement in the Virgo area, especially NGVS in the optical and GUViCS in the UV. 
We further illustrate this point in Fig. \ref{FigNgvsFuv} showing a comparison of the same area in our FUV diffuse map and in a deep $g$-band NGVS image (the brightest cirrus feature here reaches 27.2 mag arcsec$^{-2}$). This figure shows that the cirrus are also detected in these deep high quality optical images, owing to an efficient restoration of the true sky background of MegaCam data. The similar distribution of the structures at both wavelengths is convincing that it is well related to cirrus. 

\item Provide alternative extinction maps by using the correlation found between the diffuse FUV emission (as measured in the map we deliver) and the reddening, as detailed in section \ref{secdustfuv}. The 20 arcsec FUV diffuse emission is slightly better than other dust tracers based on the relation between \ebv{} and the $NUV-i$ colour of background galaxies (section \ref{sectestebv}). 
\end{itemize}

For this reason, and while there is certainly a lot of analysis and 
understanding to be still performed, we put 
these FUV maps (1 arcmin pixels based on the AIS, and 20 arcsec pixels based on deeper data, object subtracted, edited to remove artefacts, background-subtracted) on the GUViCS web site\footnote{\url{http://galex.lam.fr/guvics/mosaics.html}} and make them publicly available. We also make available the complementary maps on the same 1 arcmin pixel scale prepared and used for our analysis.

\begin{acknowledgements}

We thank the referee for interesting comments and references.
Based on observations made with the NASA Galaxy Evolution Explorer.
\GALEX{} is operated for NASA by the California Institute of Technology under NASA contract NAS5-98034.
We wish to thank the GALEX Time Allocation Committee for the generous allocation of time devoted to GUViCS.
This work is supported by the French Agence Nationale de la Recherche (ANR) Grant Program Blanc VIRAGE (ANR10-BLANC-0506-01).
This research has made use of the GOLDMine Database. 
This research has made use of the NASA/IPAC ExtraGalactic Database (NED) which is operated by the Jet Propulsion Laboratory, California Institute of Technology, under contract with the National Aeronautics and Space Administration. 
This research made use of Montage, funded by the National Aeronautics and Space Administration's Earth Science Technology Office, Computation Technologies Project, under Cooperative Agreement Number NCC5-626 between NASA and the California Institute of Technology. 
Montage is maintained by the NASA/IPAC Infrared Science Archive.
This research made use of IRAF. IRAF is distributed by the National Optical Astronomy Observatory, which is operated by the Association of Universities for Research in Astronomy (AURA) under cooperative agreement with the National Science Foundation.
C.P. acknowledges support from the Science and Technology Foundation (FCT, Portugal) through the Postdoctoral Fellowship SFRH/BPD/90559/2012, PEst-OE/FIS/UI2751/2014, and PTDC/FIS-AST/2194/2012.
S.B. thank Laurent Cambresy for Aladin tips and congratulates him for his recent HDR.

\end{acknowledgements}


\end{document}